\documentclass[11pt]{article}
\usepackage{amssymb}
\usepackage{latexsym}
\usepackage{amsmath}
\usepackage{graphicx}
\usepackage{listings}
\usepackage{cite}
\usepackage{authblk}

\usepackage{esdiff}
\usepackage{mathtools}

\def\vp{\varphi}

\def\ve{\varepsilon}
\def\al{\alpha}
\def\nb{\nabla}

\def\e{\eta}
\def\l{\lambda}
\def\k{\kappa}
\def\L{\Lambda}
\def\de{\delta}\def\om{\omega}
\title{\Large{Static black hole in minimal Horndeski gravity with Maxwell and Yang-Mills fields and some aspects of its thermodynamics}} 
\author[1,2]{\small {M. M. Stetsko\thanks{e-mail: mstetsko@gmail.com, mstetsko@upenn.edu}}}
\affil[1]{Department of Physics and Astronomy, University of Pennsylvania, Philadelphia, PA, 19104, USA}
\affil[2]{Department for Theoretical Physics, Ivan Franko National University of Lviv, Lviv, UA-79005, Ukraine}

\begin{document}
\maketitle

{\abstract{In this work we obtain a static spherically symmetric charged black hole solution in the framework of minimal Horndeski gravity with additional Maxwell and Yang-Mills fields. The obtained solution is examined, in particular its asymptotics are studied. Thermodynamics of the black hole is investigated, namely we use an effective surface gravity to derive black hole temperature. To obtain the first law of black hole thermodynamics the Wald method is applied. We also use the extended thermodynamics approach, namely it allows us to derive the Smarr relation, Gibbs free energy and the thermal equation of state. The study of thermal values in the extended space shows rich phase behaviour, in particular domain where the first order phase transition takes place and the critical point with the second order phase transition. We also study thermal behaviour near the critical point, obtain critical exponents and analyse the Ehrenfest's equations at the critical point. Finally, we calculate the Prigogine-Defay ratio confirming the conclusion about the second order phase transition at the critical point.}}

\section{Introduction}
Recent decade has become a period of outstanding progress of observational astrophysics, first of all due to long-time expected detection of the gravitational waves which required experimental setup of remarkably high accuracy\cite{Abbott_PRL16}. In general experimental observations show astonishing agreement with theoretical predictions made in the framework of General Relativity which even nowadays is exceptionally successful theory of gravity\cite{Will_LRR14}. But nonetheless on its attractive features there are some open issues which motivate people to look for alternative or more general approaches than Einsteinian theory of gravity that give answers to current puzzles. Among the most perplexing questions are the existence of singularities which as it is proved inevitably appear within general relativistic consideration, Dark Energy/Dark Matter issues, consistent description of early stage evolution of the Universe. 

To overcome the mentioned difficulties various approaches were proposed and examined giving rise to different ways of modification of general relativistic setting of the problem. The key features, their advantages and possible difficulties of the diverse approaches are given in thorough reviews \cite{Clifton_PhR12,Jimenez_PhR18,Heisenberg_PhR19,Sotiriou_RMP10,Bahamonde_RPP23}. Here we focus on Scalar-Tensor gravity theories, namely on the so-called Horndeski gravity \cite{Horndeski_IJTP74, Kobayashi_RPP19} as one of the most promising approaches. We also point out that Scalar-Tensor theories of gravity may be considered as a conservative approach, since its formulation follows the way usually used in General Relativity. We also point out that Scalar-Tensor theories have rather long history, starting back from Brans-Dicke gravity established in early 1960-ies \cite{Brans_PhysRev61}. The latter one also gained its second renaissance since the beginning of the new century, particularly because of its tight bonds with $F(R)$ gravity \cite{Sotiriou_RMP10}. Strictly speaking the Brans-Dicke theory is just a particular case of the general Horndeski gravity \cite{Kobayashi_RPP19}, but because of specific coupling between gravity and scalar sectors in Brans-Dicke-type of theories and in Horndeski gravity they are often considered separately. 

In his seminal paper \cite{Horndeski_IJTP74} Horndeski proposed the most general four dimensional Scalar-Tensor theory with the so-called derivative coupling between gravity and scalar fields which gives rise to the second order field equations. Horndeski gravity got its second revival, when relations  with the generalized Galileon model were established \cite{Kobayashi_PTP11}. The Galileons firstly appeared in studies of DGP model \cite{Dvali_PLB00}, they got their name due to a specific shift symmetry, namely $\varphi \rightarrow \varphi'=\varphi+b_{\mu}x^{\mu}+c$ ($b_{\nu}$, $c$ are constants). One of the most appealing features of Horndeski gravity related to the second order of the equations of motion is the absence of ghosts. On the other hand the Cauchy problem is well-posed in Horndeski gravity, making it an attractive model for various applications. Even though there is direct relation between the generalized Galileon theory and Horndeski gravity in four dimensions, higher dimensional generalization of Horndeski gravity has not been obtained yet \cite{Kobayashi_RPP19}. Since its relation to the DGP model and due to the fact that Horndeski theory terms in four dimensional space-time can be derived via dimensional reduction \cite{Acoleyen_PRD11,Charmousis_LNP15} it can be claimed that Horndeski gravity apart of its phenomenological origin, has some ties with String Theory, at least in the low-energy limit of the latter.  We also point out that Horndeski gravity can be generalized to become a multiscalar theory\cite{Noller_JHEP15}, other interesting generalization is the so-called DHOST theories \cite{Langlois_IJMPD19}, namely the theories with higher order equations of motion, but with some degeneracy conditions removing the Ostrogradsky instability. Horndeski gravity got numerous applications in cosmology, the most remarkable of them are pointed out in the review \cite{Kobayashi_RPP19}.

Black holes and other compact objects like neutron stars have been attracted much attention since the second revival of Horndeski gravity \cite{Rinaldi_PRD12,Minamitsuji_PRD14,Anabalon_PRD14,Cisterna_PRD14,Kobayashi_PTEP14,
Babichev_JHEP14,Bravo-Gaete_PRD14,Giribet_PRD15,Charmousis_PRD15,
Feng_JHEP15,Feng_PRD16,Stetsko_arx18,Stetsko_PRD19,Bernardo_PRD19,
Stetsko_PRD20,Baggioli_PRD21,Walia_EPJC22,BravoGaetePRD22}. Black hole solutions are important and useful toy models to study various effects, especially related to astrophysical black holes \cite{Kobayashi_PRD14}. Gravity theories including General Relativity usually have complicated structure, therefore gaining some general results valid at least within a particular gravity theory might be a problem of immense difficulty, especially for the theories beyond General Relativity. Therefore, black hole solutions are those objects which allow to derive or test implications of theory and their study is very important problem. 

Black holes in Horndeski theory are known to have nontrivial scalar field profile, particularly the scalar field may be time dependent \cite{Kobayashi_PTEP14,Kobayashi_RPP19} or/and have singular behaviour at the event horizon. Nontrivial profile of the scalar field significantly affects on various properties of the black holes and usually require careful study. Even though there a lot of black hole solutions in Horndeski gravity not much attention is paid to the case where additional fields are taken into consideration \cite{Cisterna_PRD14,Feng_PRD16,Stetsko_PRD19,Stetsko_PRD20}. It can be explained by the following two reasons, namely the first one is directly related to cumbersome structure of the Horndeski theory giving rise to equations which are hardly tractable even under quite simple assumptions. The second reason, to our mind, is related to rather general point of view that the main impact of the Horndeski gravity should take place on cosmological scales, whereas for the compact object due to various screening mechanisms, they should mimic general relativistic black holes at least for a distant observer. But studies of black holes with additional material or gauge fields in Horndeski gravity allow not only to reveal some specific features caused by the particular choice of the gravity model, they in principle may give us more general and broad view of some basic notions of black hole physics and show the range of their applicability to various gravity models.

In this paper a static black hole solution in a particular case of Horndeski theory with additional Maxwell and Yang-Mills field is considered. As far as we known the interplay of Horndeski gravity and Yang-Mills field even though both of them are taken probably in their simplest form is studied for the first time. The Maxwell field in its standard form as well as for some of its nonlinear generalizations were considered in case of Horndeski theory \cite{Cisterna_PRD14,Feng_PRD16,Stetsko_PRD19,Stetsko_PRD20}, whereas nonabelian fields were examined mainly within General Relativity \cite{Yasskin_PRD75,Kasuya_PRD82,Lavrelashvili_NPB93,Tachizawa_PRD95,Volkov_PRep99,
Brihaye_PRD07,Mazhari_PRD07,Mazhari_PLB08,Brihaye_PRD11,Ghosh_PLB11}, or more general in Einstein-dilaton theory \cite{Lavrelashvili_NPB93,Stetsko_PRD20_2,Stetsko_GRG21,Stetsko_IJMPA21}. We also take into account the Maxwell field to examine interplay between the gauge fields in the framework of Horndeski gravity and as we will show there is an effective ``coupling" between them which does not appear neither in General Relativity, nor in a more general Einstein-dilaton theory \cite{Stetsko_GRG21,Stetsko_IJMPA21}. We also pay considerable attention to study of various aspects of thermodynamics for the obtained solution.

The work is organized as follows. In the following section we obtain and study a static black hole solution in Horndeski gravity with additional abelian and nonabelian gauge fields. In the third section we obtain and examine black hole temperature. In the section four we use Wald approach to derive the first law of black hole thermodynamics, obtain other thermodynamics values such as entropy and heat capacity and examine the latter one. In the fifth section we use the extended thermodynamics approach to derive the extended first law and the Smarr relation. In the sixth section we obtain the Gibbs free energy and study its behaviour. Critical behaviour in the extended approach is studied in the section seven. Finally, in the last section there are some conclusions and future prospects.

\section{Equations of motion for the theory with nonminimal derivative coupling and static black hole's solution}
General Horndeski gravity gives rise to complicated equations  which even for the geometries with high symmetry are difficult to handle with, therefore we consider one of its simplest particular cases, but which inherits distinctive feature of the general Horndeski gravity, namely its specific derivative coupling between gravity and additional scalar field. Similarly to general Horndeski gravity the equations of motion are of the second order making the theory free from Ostrogradski instability. We also take into account some gauge fields, namely we consider both abelian (electromagnetic) and nonabelian ones which are minimally coupled to gravity. The action for our system can be written in the form: 
 \begin{equation}\label{action}
S=\frac{1}{16\pi}\int d^{n+1}x\sqrt{-g}\left( R-2\Lambda-\frac{1}{2}\left(\alpha g^{\mu\nu}-\eta G^{\mu\nu}\right)\partial_{\mu}\vp\partial_{\nu}\vp -{\rm Tr}(F^{(a)}_{\mu\nu}F^{(a)\mu\nu})-\cal{F}_{\mu\nu}\cal{F}^{\mu\nu}\right)+S_{GHY},
\end{equation}
where  $g_{\mu\nu}$ and $g$ is the metric tensor and its determinant respectively, $R$ and $G_{\mu\nu}$ are the Ricci scalar and the Einstein tensor correspondingly, $\vp$ is the scalar field and $\al$  and $\eta$ are coupling constants for it and finally $F^{(a)}_{\mu\nu}$ and ${\cal F}_{\mu\nu}$ are field strengths for nonabelian and ablelian fields respectively. We note that since there is no potential for the scalar field in the action (\ref{action}), functions we obtain and analyse show their dependence on the ratio of the coupling parameters $\al/\eta$, therefore only one of them can be treated as a parameter which can be varied, but here we keep both in order to consider some limit cases. We also point out that $S_{GHY}$ term in the action (\ref{action}) denotes the so called boundary Gibbons-Hawking-York  term which makes the variational problem well-defined. For this theory with nonminimal derivative coupling the Gibbons-Hawking-York term can be written in the form:
\begin{equation}\label{GHY_nm}
S_{GHY}=\frac{1}{8\pi}\int d^nx\sqrt{|h|}\left(K+\frac{\e}{4}\left[\nb^{\mu}\vp\nb^{\nu}\vp K_{\mu\nu}+(n^{\mu}n^{\nu}\nb_{\mu}\vp \nb_{\nu}\vp+(\nb\vp)^2)K\right]\right),
\end{equation}
where $h$ is the determinant of the boundary metric $h_{\mu\nu}$, $K_{\mu\nu}$ and $K$ denote the extrinsic curvature tensor and its trace correspondingly and finally $n^{\mu}$ is the vector normal to the boundary hypersurface.

We point out here that the field tensors for the gauge fields are defined in the standard way, namely for the Yang-Mills field we write:
\begin{equation}\label{YM_field}
F^{(a)}_{\mu\nu}=\partial_{\mu}A^{(a)}_{\nu}-\partial_{\nu}A^{(a)}_{\mu}+\frac{1}{\bar{\sigma}}C^{(a)}_{(b)(c)}A^{(b)}_{\mu}A^{(c)}_{\nu},
\end{equation} 
where $A^{(a)}_{\mu}$ is the Yang-Mills potential, $\bar{\sigma}$ is the coupling constant for nonabelian field and $C^{(a)}_{(b)(c)}$ are the structure constants for corresponding gauge group. In this work the gauge group is chosen to be the special orthogonal one $SO(n)$. 

The Maxwell field tensor is defined in the standard fashion:
\begin{equation}\label{Max_field}
{\cal{F}}_{\mu\nu}=\partial_{\mu}{\cal{A}}_{\nu}-\partial_{\nu}{\cal{A}}_{\mu},
\end{equation}
and here ${\cal A}_{\mu}$ is the Maxwell field potential.

To obtain equations of motion for the system given by the action (\ref{action}) the least action principle is used. For gravitational part we can write:
\begin{equation}\label{einstein}
{\cal E}_{\mu\nu}:=G_{\mu\nu}+\Lambda g_{\mu\nu}-\left(\frac{1}{2}(\alpha T^{(1)}_{\mu\nu}+\eta T^{(2)}_{\mu\nu})+T^{(3)}_{\mu\nu}+T^{(4)}_{\mu\nu}\right)=0,
\end{equation}
where we have:
\begin{equation}\label{scal_min}
T^{(1)}_{\mu\nu}=\nb_{\mu}\vp\nb_{\nu}\vp-\frac{1}{2}g_{\mu\nu}\nb^{\lambda}\vp\nb_{\lambda}\vp,
\end{equation}
\begin{eqnarray}\label{scal_nm}
\nonumber T^{(2)}_{\mu\nu}=\frac{1}{2}\nb_{\mu}\vp\nb_{\nu}\vp R-2\nb^{\lambda}\vp\nb_{\nu}\vp R_{\lambda\mu}+\frac{1}{2}\nb^{\lambda}\vp\nb_{\lambda}\vp G_{\mu\nu}-g_{\mu\nu}\left(-\frac{1}{2}\nb_{\lambda}\nb_{\kappa}\vp\nb^{\lambda}\nb^{\kappa}\vp\right.\\\left.+\frac{1}{2}(\nb^2\vp)^2-R_{\lambda\kappa}\nb^{\lambda}\vp\nb^{\kappa}\vp\right)
-\nb_{\mu}\nb^{\lambda}\vp\nb_{\nu}\nb_{\lambda}\vp+
\nb_{\mu}\nb_{\nu}\vp\nb^2\vp-R_{\lambda\mu\kappa\nu}\nb^{\lambda}\vp\nb^{\kappa}\vp,
\end{eqnarray}
\begin{equation}\label{YM_str}
T^{(3)}_{\mu\nu}=2{\rm Tr}\left(F^{(a)}_{\mu\l}{F^{(a)\l}_{\nu}}\right)-\frac{g_{\mu\nu}}{2}{\rm Tr}\left(F^{(a)}_{\l\k}{F^{(a)\l\k}}\right),
\end{equation}
\begin{equation}\label{max_str}
T^{(4)}_{\mu\nu}=2{\cal F}_{\mu\l}{{\cal F}_{\nu}}^{\l}-\frac{g_{\mu\nu}}{2}{\cal F}_{\l\k}{\cal F}^{\l\k}.
\end{equation}
It is clear that in the right hand side of the equation (\ref{einstein}) there are stress-energy tensors for the scalar and gauge fields given by the upper relations (\ref{scal_min})-(\ref{max_str}).

The least action principle also allows us to obtain equations of motion for the scalar and the gauge fields. For the scalar field $\vp$ we arrive at the   following equation:
\begin{equation}\label{scal_f_eq}
{\cal E}_{\vp}:=(\alpha g_{\mu\nu}-\eta G_{\mu\nu})\nb^{\mu}\nb^{\nu}\vp=0.
\end{equation}
For the Yang-Mills field we obtain:
\begin{equation}\label{YM_eq}
{{\cal E}_A}^{(a)\nu}:=\nabla_{\mu}(F^{(a)\mu\nu})+\frac{1}{\bar{\sigma}}C^{(a)}_{(b)(c)}A^{(b)}_{\mu}F^{(c)\mu\nu}=0.
\end{equation}
Finally, for the abelian gauge field the standard Maxwell equations can be derived:
\begin{equation}\label{em_eq}
{{\cal E}_{\cal A}}^{\nu}:=\nabla_{\mu}{\cal{F}}^{\mu\nu}=0.
\end{equation}

Here we are going to obtain a static black hole's solution therefore the general form of the metric can be written in the  following form:
\begin{equation}\label{metric}
ds^2=-U(r)dt^2+W(r)dr^2+r^2d\Omega^{2}_{(n-1)},
\end{equation}
where $\Omega^{2}_{(n-1)}$ represents the element of length for a unit $n-1$--dimensional hypersphere and the metric functions $U(r)$
and $W(r)$ will be obtained from the equations of motion. We also point out here that in the present work we assume that $n\geqslant 3$.

For a static electrically charged solution the gauge potential for the Maxwell (abelian) field can be chosen in the form ${\cal A}={\cal A}_0(r)dt$. From the Maxwell equations (\ref{em_eq}) we derive immediately that the electromagnetic field takes the form:
\begin{equation}\label{Max_field}
{\cal F}_{rt}=\frac{q}{r^{n-1}}\sqrt{UW}
\end{equation}

It is known that the so-called Wu-Yang ansatz \cite{Yasskin_PRD75,Lavrelashvili_NPB93,Brihaye_PRD07,Mazhari_PRD07,Stetsko_PRD20_2} being one of the simplest possible choices to satisfy the Yang-Mills equations (\ref{YM_eq}) allowed to derive various solutions in pure Yang-Mills theory and if gravity was taken into account it brought to nontrivial black hole solutions. Therefore, the nonabelian gauge potential takes the form as follows:
\begin{equation}\label{YM_pot}
{\bf A}^{(a)}=\frac{\bar{q}}{r^2}C^{(a)}_{(i)(j)}x^{i}dx^{j}, \quad r^2=\sum^{n}_{j=1}x^2_j,
\end{equation}
here we point out that in order to satisfy the equations of motion (\ref{YM_eq}) we impose that $\bar{q}=\bar{\sigma}$ and for simplicity auxiliary Cartesian coordinates $x^i$ were used and the indices $a, i, j$ can take the following values: $1\leqslant a \leqslant n(n-1)/2$, $2\leqslant j+1<i\leqslant n$. The relations between the coordinates $x^i$ and the angular variables of a spherical coordinate system are standard:
\begin{eqnarray}
\nonumber x_1=r\cos{\chi_{n-1}}\sin{\chi_{n-2}}\ldots\sin{\chi_1},\quad x_2=r\sin{\chi_{n-1}}\sin{\chi_{n-2}}\ldots\sin{\chi_1},\\
\nonumber  x_3=r\cos{\chi_{n-2}}\sin{\chi_{n-3}}\ldots\sin{\chi_1},\quad x_4=r\sin{\chi_{n-2}}\sin{\chi_{n-3}}\ldots\sin{\chi_1},\\
\nonumber \cdots \quad\quad\\
x_n=r\cos{\chi_1},
\end{eqnarray}
and the angular variables $\chi_i$ have typical ranges of variation, namely for $1\leqslant i\leqslant n-2$ we have $0\leqslant\chi_i\leqslant\pi$ and $0\leqslant\chi_{n-1}<2\pi$. Using the angular variables we can also represent the length element for the unit sphere:
\begin{equation}
d\Omega^2_{n-1}=d\chi^2_{1}+\sum^{n-1}_{j=2}\prod^{j-1}_{i=1}\sin^2{\chi_{i}}d\chi^2_{j}.
\end{equation}
The gauge potential (\ref{YM_pot}) might be rewritten in terms of angular variables, but its explicit form would not be as simple as for the Cartesian ones. Using the relation (\ref{YM_field}) one can calculate the gauge field $F^{(a)}_{\mu\nu}$ and check that the equations of motion (\ref{YM_eq}) are satisfied. The invariant for the Yang-Mills potential can be calculated, namely we arrive at:
\begin{equation}
{\rm Tr}(F^{(a)}_{\rho\sigma}F^{(a)\rho\sigma})=(n-1)(n-2)\frac{\bar{q}^2}{r^4}.
\end{equation}

Using the metric ansatz (\ref{metric}) and taking into account the gauge field tensors and their invariants we can write the equations (\ref{einstein}) in the following form:
\begin{gather}
\nonumber\frac{(n-1)}{2rW}\left(\frac{W'}{W}+\frac{(n-2)}{r}(W-1)\right)\left(1+\frac{3}{4}\e\frac{(\vp')^2}{W}\right)-\L=\frac{\al}{4W}(\vp')^2+\\\frac{\e}{2}\left(\frac{(n-1)(n-2)}{r^2W^2}\left(W-\frac{1}{2}\right)(\vp')^2+\frac{(n-1)}{rW^2}\vp''\vp'\right)+\frac{q^2}{r^{2(n-1)}}+\frac{(n-1)(n-2)\bar{q}^2}{2r^4},\label{eins_00}
\end{gather}
\begin{gather}
\nonumber\frac{(n-1)}{2rW}\left(\frac{U'}{U}-\frac{(n-2)}{r}(W-1)\right)\left(1+\frac{3}{4}\e\frac{(\vp')^2}{W}\right)+\L=\\\frac{\al}{4W}(\vp')^2-\frac{\e(n-1)(n-2)}{4r^2W}(\vp')^2-\frac{q^2}{r^{2(n-1)}}-\frac{(n-1)(n-2)\bar{q}^2}{2r^4},\label{eins_11}
\end{gather}
where prime denotes the derivative with respect to $r$.

The equation for the scalar field (\ref{scal_f_eq}) may be also integrated at least for a once, as a result we obtain:
\begin{equation} 
\sqrt{\frac{U}{W}}r^{n-1}\left[\al-\e\frac{(n-1)}{2rW}\left(\frac{U\rq{}}{U}-\frac{(n-2)}{r}(W-1)\right)\right]\vp'=C,
\end{equation}
where $C$ is an integration constant. The latter relation might be used to express the derivative $\vp'$ in terms of the metric functions, their derivatives and also as an explicit function of the radius $r$, but in this general case the corresponding relation for the function $\vp'$ turns to be of a complicated form and it would be difficult to operate with it. Therefore, for simplicity of the following calculations we set $C=0$, even though the condition we impose gives rise just to a particular solution, but it is quite nontrivial and it is worth being studied. The condition $C=0$ is equivalent to the following constraint:
\begin{equation}\label{cond_1}
\al g_{rr}-\e G_{rr}=0.
\end{equation}
Here we point out that the same condition (\ref{cond_1}) was used in our earlier works \cite{Stetsko_PRD19,Stetsko_PRD20} as well as in papers of other author were black holes with nonminimal derivative coupling were studied \cite{Rinaldi_PRD12,Anabalon_PRD14,Minamitsuji_PRD14}.

Now the equations (\ref{eins_00})-(\ref{eins_11}) can be solved together with the upper relation (\ref{cond_1}). As a result we obtain:
\begin{equation}\label{vp_prime}
\left(\vp'\right)^2=-\frac{4r^2 W}{2\al r^2+\e(n-1)(n-2)}\left(\L+\frac{\al}{\e}+q^2r^{2(1-n)}+\frac{(n-1)(n-2)}{2}\bar{q}^2r^{-4}\right);
\end{equation}
\begin{equation}\label{UW_prod}
UW=\frac{\left((\al-\L\e)r^2+\e(n-1)(n-2)-\e q^2r^{2(2-n)}-\e(n-1)(n-2)\bar{q}^2r^{-2}/2\right)^2}{(2\al r^2+\e(n-1)(n-2))^2}.
\end{equation}
The square of the derivative $\vp'$ has to be positive outside of the black hole, it might be achieved if some conditions on the parameters $\al$, $\e$, $\L$, $q$  and $\bar{q}$ are imposed. For instance, when both parameters $\al$ and $\e$ are positive, the cosmological constant $\L$ should be negative to provide positivity of the $(\vp')^2$  in the outer domain. Similar conclusion is inferred if we impose $\al>0$ and $\e<0$.

Finally, the metric function $U(r)$ can be written in the following form:
 \begin{gather}
\nonumber U(r)=1-\frac{\mu}{r^{n-2}}-\frac{2\L}{n(n-1)}r^{2}-\frac{(n-2)}{(n-4)}\frac{\bar{q}^2}{r^2}+\frac{2q^{2}}{(n-1)(n-2)}r^{2(2-n)}+\nonumber\frac{1}{2\al\e(n-1)r^{n-2}}\times\\\nonumber\left((\al+\L\e)^2\int\frac{r^{n+1}}{r^2+d^2}dr+\e^2q^4\int\frac{r^{5-3n}}{r^2+d^2}dr+2\e (\al+\L\e)q^2\int\frac{r^{3-n}}{r^2+d^2}dr+(n-1)(n-2)\times\right.\\\left.\e\bar{q}^2\left((\al+\L\e)\int\frac{r^{n-3}}{r^2+d^2}dr+\e q^2\int\frac{r^{-(n+1)}}{r^2+d^2}dr+\frac{\e}{4}(n-1)(n-2)\bar{q}^2\int\frac{r^{n-7}}{r^2+d^2}dr\right)\right),\label{U_mf}
 \end{gather}
 where $d^2=\e(n-1)(n-2)/2\al$. In the upper relation indefinite integrals are used, because of some peculiarities for even and odd dimensions of space $n$. We also point out that the fourth term in the function $U(r)$ (\ref{U_mf}) has additional logarithmic factor ($\sim\ln{r}$) if $n=4$.

To obtain the explicit form of the metric function (\ref{U_mf}) all the integrals in (\ref{U_mf}) should be rewritten in terms of corresponding functions. Due to some subtleties for even and odd dimensions of space we write:
\begin{equation}\label{int_1}
\int\frac{r^m}{r^2+d^2}dr=\sum^{(m-2)/2}_{j=0}(-1)^jd^{2j}\frac{r^{m-2j-1}}{m-2j-1}+(-1)^{\frac{m}{2}}d^{m-1}\arctan\left(\frac{r}{d}\right),
\end{equation} 
and here $m$ is a positive even number.  If $m$ is a positive odd number  the latter integral might be written in the form:
\begin{equation}\label{int_2}
\int\frac{r^m}{r^2+d^2}dr=\sum^{(m-3)/2}_{j=0}(-1)^jd^{2j}\frac{r^{m-2j-1}}{m-2j-1}+(-1)^{\frac{m-1}{2}}\frac{d^{m-1}}{2}\ln\left(1+\frac{r^2}{d^2}\right),
\end{equation} 
and if in the latter relation $m=1$ there is just a logarithmic contribution.  There are also integrals of the form:
\begin{equation}\label{int_3}
\int\frac{r^{-m}}{r^2+d^2}dr=\sum^{(m-2)/2}_{j=0}(-1)^j\frac{r^{1+2j-m}}{(1+2j-m)d^{2(j+1)}}+\frac{(-1)^{\frac{m}{2}}}{d^{m+1}}\arctan\left(\frac{r}{d}\right),
\end{equation}
where $m$ is an even positive. While in case of an odd integer the upper integral takes the form as follows:
 \begin{equation}\label{int_4}
\int\frac{r^{-m}}{r^2+d^2}dr=\sum^{(m-3)/2}_{j=0}(-1)^j\frac{r^{1+2j-m}}{(1+2j-m)d^{2(j+1)}}+\frac{(-1)^{\frac{m+1}{2}}}{2d^{m+1}}\ln\left(1+\frac{d^2}{r^2}\right).
\end{equation}
Having used the written above integrals we write the explicit form for the metric function $U(r)$ (\ref{U_mf}). In particular, for odd $n$ we obtain:
\begin{gather}
\nonumber U(r)=1-\frac{\mu}{r^{n-2}}-\frac{2\L}{n(n-1)}r^{2}-\frac{(n-2)}{(n-4)}\frac{\bar{q}^2}{r^2}+\frac{2q^{2}}{(n-1)(n-2)}r^{2(2-n)}+\nonumber\frac{1}{2\al\e(n-1)}\times\\\nonumber\left[(\al+\L\e)^2\left(\sum^{\frac{n-1}{2}}_{j=0}(-1)^jd^{2j}\frac{r^{2(1-j)}}{n-2j}+(-1)^{\frac{n+1}{2}}\frac{d^n}{r^{n-2}}\arctan\left(\frac{r}{d}\right)\right)+2\e(\al+\L\e)q^2\left(\sum^{\frac{n-5}{2}}_{j=0}\frac{(-1)^{j}r^{6-2n+2j}}{(4-n+2j)d^{2(j+1)}}+\right.\right.\\\left.\left.\nonumber\frac{(-1)^{\frac{n-3}{2}}}{d^{n-2}r^{n-2}}\arctan\left(\frac{r}{d}\right)\right)+\e^2q^4\left(\sum^{\frac{3n-7}{2}}_{j=0}\frac{(-1)^{j}r^{2(4+j-2n)}}{(6+2j-3n)d^{2(j+1)}}+\frac{(-1)^{\frac{3n-5}{2}}}{d^{3n-4}r^{n-2}}\arctan\left(\frac{r}{d}\right)\right)+\e(n-1)(n-2)\bar{q}^2\times\right.\\\left.\nonumber\left((\al+\L\e)\left(\sum^{\frac{n-5}{2}}_{j=0}\frac{(-1)^{j}d^{2j}r^{-2(1+j)}}{n-4-2j}+(-1)^{\frac{n-3}{2}}\frac{d^{n-4}}{r^{n-2}}\arctan\left(\frac{r}{d}\right)\right)+\e q^2\left(\sum^{\frac{n-1}{2}}_{j=0}\frac{(-1)^{j}r^{2(1+j-n)}}{(2j-n)d^{2(j+1)}}+\right.\right.\right.\\\left.\left.\left.+\frac{(-1)^{\frac{n+1}{2}}}{d^{n+2}r^{n-2}}\arctan\left(\frac{r}{d}\right)\right)+\e(n-1)(n-2)\frac{\bar{q}^2}{4}\left(\sum^{\frac{5-n}{2}}_{j=0}\frac{(-1)^jr^{2(j-2)}}{(2j+n-6)d^{2(j+1)}}+\frac{(-1)^{\frac{7-n}{2}}}{d^{8-n}r^{n-2}}\arctan\left(\frac{r}{d}\right)\right)\right)\right].\label{U_odd_1}
\end{gather} 
It should be pointed that the above relation, namely its last sum is valid when $n<7$. If $n>7$ in the last integral in the relation (\ref{U_mf}) we will have $n-7>0$, therefore instead of the relation (\ref{int_3}) one should use the relation (\ref{int_1}).

For even $n$ we write:
\begin{gather}
\nonumber U(r)=1-\frac{\mu}{r^{n-2}}-\frac{2\L}{n(n-1)}r^{2}-\frac{(n-2)}{(n-4)}\frac{\bar{q}^2}{r^2}+\frac{2q^{2}}{(n-1)(n-2)}r^{2(2-n)}+\nonumber\frac{1}{2\al\e(n-1)}\times\\\nonumber\left[(\al+\L\e)^2\left(\sum^{\frac{n-2}{2}}_{j=0}(-1)^jd^{2j}\frac{r^{2(1-j)}}{n-2j}+(-1)^{\frac{n}{2}}\frac{d^n}{2r^{n-2}}\ln\left(1+\frac{r^2}{d^2}\right)\right)+2\e(\al+\L\e)q^2\left(\sum^{\frac{n-6}{2}}_{j=0}\frac{(-1)^{j}r^{6-2n+2j}}{(4-n+2j)d^{2(j+1)}}+\right.\right.\\\left.\left.\nonumber\frac{(-1)^{\frac{n-2}{2}}}{2(dr)^{n-2}}\ln\left(1+\frac{d^2}{r^2}\right)\right)+\e^2q^4\left(\sum^{\frac{3n-8}{2}}_{j=0}\frac{(-1)^{j}r^{2(4+j-2n)}}{(6+2j-3n)d^{2(j+1)}}+\frac{(-1)^{\frac{3n-4}{2}}}{2d^{3n-4}r^{n-2}}\ln\left(1+\frac{d^2}{r^2}\right)\right)+\e(n-1)(n-2)\bar{q}^2\times\right.\\\left.\nonumber\left((\al+\L\e)\left(\sum^{\frac{n-6}{2}}_{j=0}\frac{(-1)^{j}d^{2j}r^{-2(1+j)}}{n-4-2j}+(-1)^{\frac{n-2}{2}}\frac{d^{n-4}}{2r^{n-2}}\ln\left(1+\frac{r^2}{d^2}\right)\right)+\e q^2\left(\sum^{\frac{n-2}{2}}_{j=0}\frac{(-1)^{j}r^{2(1+j-n)}}{(2j-n)d^{2(j+1)}}+\right.\right.\right.\\\left.\left.\left.+\frac{(-1)^{\frac{n+2}{2}}}{2d^{n+2}r^{n-2}}\ln\left(1+\frac{d^2}{r^2}\right)\right)+\e(n-1)(n-2)\frac{\bar{q}^2}{4}\left(\sum^{\frac{n-10}{2}}_{j=0}\frac{(-1)^jd^{2j}r^{-2(j+3)}}{n-8-2j}+(-1)^{\frac{n-8}{2}}\frac{d^{n-8}}{2r^{n-2}}\ln\left(1+\frac{r^2}{d^2}\right)\right)\right)\right].\label{U_even_1}
\end{gather}
It should be emphasized that in contrast with the relation (\ref{U_odd_1}) the upper relation is written for the case $n>7$. If $n<7$ for even $n$ to calculate the last integral in the relation (\ref{U_mf}) one should use the relation (\ref{int_4}) instead of (\ref{int_2}) which is taken in (\ref{U_even_1}). Due to special interest to lower dimensions, we also write explicit forms for metric functions when $n=3$ and $n=4$. Namely, if $n=3$ the metric function $U(r)$ takes the following form:
\begin{gather}
\nonumber U(r)=1-\frac{\mu}{r}-\frac{\L}{3}r^2+\frac{q^2+\bar{q}^2}{r^2}+\frac{1}{4\al\e}\left[(\al+\L\e)^2\left(\frac{r^2}{3}-d^2\right)+\frac{\e^2(q^2+\bar{q}^2)^2}{d^2r^2}\times\right.\\\left.\left(\frac{1}{d^2}-\frac{1}{3r^2}\right)+\left((\al+\L\e)^2d+\frac{\e(q^2+\bar{q}^2)}{d^3}\right)^2\frac{d}{r}\arctan\left(\frac{r}{d}\right)\right].\label{U_n3}
\end{gather}
Here we would like to point out that in the three-dimensional case ($n=3$) the contribution of abelian electric and noanbelian magnetic fields are completely identical to each other. For the case $n=4$ the metric function $U(r)$ is as follows:
\begin{gather}
\nonumber U(r)=1-\frac{\mu}{r^2}-\frac{\L}{6}r^2+\frac{q^2}{3r^4}-2\frac{\bar{q}^2}{r^2}\ln{\left(\frac{r}{d}\right)}+\frac{1}{6\al\e}\left[\frac{(\al+\L\e)^2}{2}\left(\frac{r^2}{2}-d^2\right)+\left((\al+\L\e)\frac{d^4}{2}+3\e\bar{q}^2\right)\times\right.\\\left.\nonumber\frac{(\al+\L\e)}{r^2}\ln\left(1+\frac{r^2}{d^2}\right)+\frac{\e^2q^4}{2d^2r^4}\left(-\frac{1}{3r^4}+\frac{1}{2(dr)^2}-\frac{1}{d^4}\right)+3\frac{\e^2q^2\bar{q}^2}{d^2r^4}\left(\frac{1}{d^2}-\frac{1}{2r^2}\right)-\frac{9\e^2\bar{q}^4}{2d^2r^4}+\right.\\\left.\left(\frac{\e^2}{2d^2}\left(\frac{q^2}{d^2}-3\bar{q}^2\right)^2-\e(\al+\L\e)q^2\right)\frac{1}{d^2r^2}\ln\left(1+\frac{d^2}{r^2}\right)\right].\label{U_n4}
\end{gather}
\begin{figure}
\centerline{\includegraphics[scale=0.33,clip]{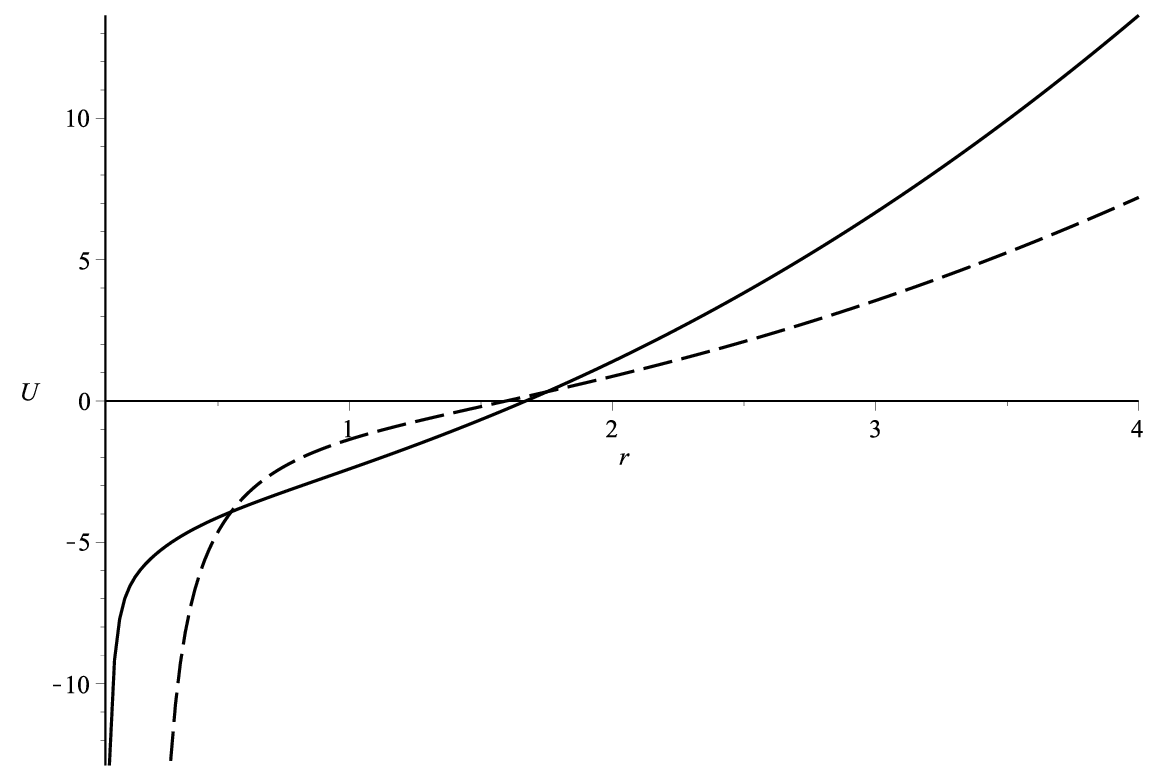}\includegraphics[scale=0.33,clip]{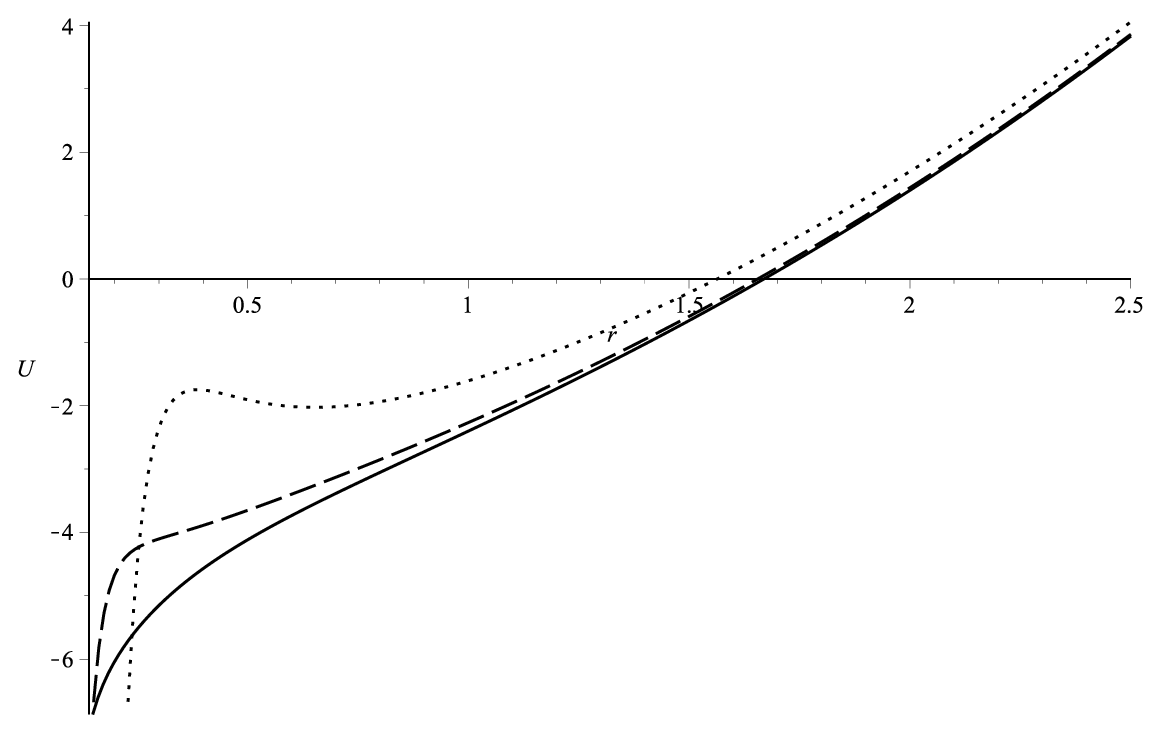}}
\caption{Metric functions $U(r)$ for various dimensions $n$ (the left graph) and different values of the electric charge $q$ (the right graph). For both graphs we have $\L=-2$, $\al=0.2$, $\e=0.4$, $\bar{q}=0.2$. For the left graph we have $q=0.2$ and solid and dashed lines correspond to $n=3$ and $n=4$ respectively. For the right graph we have taken $n=3$ and solid, dashed and dotted curves correspond to $q=0.2$, $q=0.4$ and $q=0.8$ respectively.}\label{metr_f_gr}
\end{figure}
Even though the explicit expressions for the metric functions (\ref{U_odd_1}) and (\ref{U_even_1}), as well as their particular cases (\ref{U_n3}) and (\ref{U_n4}) correspondingly, are rather cumbersome, some important conclusions about their behaviour can be derived relatively easily. First, for both types of parity of dimensions the behaviour of the metric function $U(r)$ for large distances is asymptotically of AdS-type if both coupling parameters $\al$ and $\e$ are positive (of the same sign), namely we can write:
\begin{equation}\label{U_as_inf}
U\simeq \frac{(\al-\L\e)^2}{2n(n-1)\al\e}r^2=\frac{\eta\left(\al/\e-\L\right)^2}{2n(n-1)\al}r^2.
\end{equation} 
Since the gauge fields give decaying terms in outer far zone, it is natural that there is anti-de Sitterian term which shows leading behaviour if $r\to\infty$, similar results were obtained if nonlinear electromagnetic field was taken into account  \cite{Stetsko_PRD19, Stetsko_PRD20}.

If the radius $r$ becomes very small ($r\rightarrow 0$) the metric function $U(r)$ shows singular behaviour which is defined by electromagnetic field part in general case. Namely, when $r\rightarrow 0$ we can write:
\begin{equation}\label{as_n_r0}
U(r)\simeq -\frac{q^4}{3(n-1)^2(n-2)^2}r^{4(2-n)}.
\end{equation}
Therefore, the leading term for $r\to 0$ is related just to the Maxwell field what is rather expectable, but we point out that this asymptotic appears, because of interplay of the gauge field term with a specific Horndeski theory influence, even though the asymptotic (\ref{as_n_r0}) does not show explicit dependence on coupling parameters $\al$ and $\eta$. We point out that the Yang-Mills terms alone or the terms where effective coupling between the Maxwell and Yang-Mills fields are taken into account,  show less singular behaviour in comparison with term (\ref{as_n_r0}) in the limit $r\to 0$ if $n>3$. For $n=3$ both gauge fields give rise to contributions of the same order what is clearly seen from the explicit form of the metric function for this case (\ref{U_n3}). Namely, in this case we have:
\begin{equation}\label{as_n3_r0}
U(r)\simeq -\frac{(q^2+\bar{q}^2)^2}{12 r^4}.
\end{equation} 
One of the most important conclusions following from the asymptotic expressions (\ref{as_n_r0}) and (\ref{as_n3_r0}) because of their negative signs and it means that the singular behaviour of the metric function $U(r)$ if $r\to 0$ is more similar to the Schwarzschild black hole than to the Reissner-Nordstrom one as it might be expected. That character of behaviour of the metric function is also clearly reflected on the graph of the metric function $U(r)$ given by the Figure [\ref{metr_f_gr}]. The second of the graphs of the Figure [\ref{metr_f_gr}] also implies that apart of the only event horizon, namely the point where the function $U(r)$ crosses the horizontal axis additional inner horizons may appear if one increases the electric charge $q$, but it also may occur if the parameter $\bar{q}$ goes up, but a detailed analysis of this issue will be considered elsewhere. The other important conclusion which is also directly related to the mentioned above features is that for any charge $q$ or $\bar{q}$ a naked singularity never occurs as it usually takes place within General Relativity if the charge of a black hole increases while other parameters of the black hole are held fixed.

We also briefly examine the particular case if $\al=0$, namely when only derivative  coupling between gravity and scalar field part is considered. The particular solution for $\al=0$ is substantially simpler than the general one examined above. Namely, for the squared derivative $(\vp')^2$  and the product of the metric functions we obtain:
\begin{gather}
(\vp')^2=\frac{4r^2W}{\e(n-1)(n-2)}\left(\L+q^2r^{2(1-n)}+\frac{1}{2}(n-1)(n-2)\bar{q}^2r^{-4}\right),\\
UW=\left(1-\frac{q^2}{(n-1)(n-2)}r^{2(2-n)}-\frac{\bar{q}^2}{2r^2}\right)^2.
\end{gather}
The metric function $U(r)$ can be written in the form:
\begin{multline}\label{U_al0}
U(r)=1-\frac{\mu}{r^{n-2}}-\frac{2\L}{(n-1)(n-2)}r^2-\frac{(n-2)}{(n-4)}\frac{\bar{q}^2}{r^2}+\frac{2q^2}{(n-1)(n-2)}r^{2(2-n)}+\\\frac{\L^2}{(n-1)^2(n^2-4)}r^4-\frac{q^4}{3(n-1)^2(n-2)^2}r^{4(2-n)}-\frac{2\L q^2}{(n-1)^2(n-2)(n-4)}r^{2(3-n)}\\+\frac{\L\bar{q}^2}{(n-1)(n-2)}-\frac{q^2\bar{q}^2}{n(n-1)}r^{2(1-n)}+\frac{(n-2)}{4(n-6)}\frac{\bar{q}^4}{r^4}.
\end{multline}
It should be emphasized that in (\ref{U_al0}) we impose $n\neq 4$ and $n\neq 6$, if for instance $n=4$ the forth term in the upper row and the third term in the middle one take additional logarithmic factor ($\sim\ln{r}$) and if $n=6$ this factor appears in the last term in the bottom row, but for both cases it does not change drastically qualitative behaviour of the metric function $U(r)$. We would like to note that for the particular case $\al=0$ neither the product $UW$ nor the function $U(r)$ depend on the parameter $\e$. We point out that if $r\to\infty$ the leading term of the metric function is of the order $\sim\L^2r^4$, and it is suppressed if $\al\neq  0$, since this term is always positive it gives rise to the conclusion that there is no cosmological horizon for any sign of the cosmological constant. If $r\to 0$ the leading term of the metric (\ref{U_al0}) is the same as for the general case, namely (\ref{as_n_r0}) and to some extent it is expectable since for small distances the metric is mainly defined by the leading electromagnetic field term. We also note that the product $UW\to 1$ if $r\to\infty$ and it becomes singular if $r\to 0$, but this singular behaviour, which also takes place if $\al\neq 0$ allows to moderate singularities for the invariants of the Riemann tensor in comparison with standard general relativity solutions \cite{Stetsko_PRD19}.  
\section{Black hole temperature}
One of the basic notions of black holes thermodynamics is temperature. The definition of the temperature is based on geometrical notion of surface gravity which can be applied not only to black holes within General Relativity, but also to more general gravitational frameworks \cite{Racz_CQG96,Sarkar_PRD13,Ghosh_PRD20} including Horndeski gravity \cite{Sang_PRD21}.  The surface gravity $\kappa$ is defined as  follows:
\begin{equation}\label{surf_grav}
\kappa^2=-\frac{1}{2}\nabla_{a}\chi_b\nabla^{a}\chi^{b},
\end{equation}
where $\chi^{\mu}$ is a Killing vector, which is null on the event horizon. Since in our work the static configuration (\ref{metric}) is considered, the time translation vector $\chi^{\mu}=\partial/\partial t$ satisfies the mentioned condition. In the framework of General Relativity and in various other approaches to gravity the temperature is defined to be proportional to the surface gravity, namely:
\begin{equation}
T_{BH}=\frac{\kappa}{2\pi}=\frac{1}{4\pi}\frac{U'(r_+)}{\sqrt{U(r_+)W(r_+)}},
\end{equation} 
where $r_+$ denotes the event horizon of the black hole. Having calculated the derivative $U'(r_+)$ and after simple algebra we write the temperature in the form:
\begin{equation}\label{T_aux}
T_{BH}=\frac{1}{4\pi(n-1)r_{+}}\left(\left(\frac{\al}{\eta}-\L\right)r^2_{+}+(n-1)(n-2)-\frac{q^2}{r^{2(n-2)}_{+}}-\frac{(n-1)(n-2)}{2r^2_{+}}\bar{q}^2\right).
\end{equation}
Surface gravity has clear geometric meaning and as it is mentioned above it is widely applicable, including Horndeski theory \cite{Sang_PRD21}, but even for the latter theory there are some subtleties. The authors \cite{Sang_PRD21} consider particular case of general Horndeski gravity similar to that considered here, but they also make several assumptions which single out a particular class of solutions that can be easily reduced to some general relativistic ones if Horndeski coupling parameter $\e$ is turned off. It is also supposed that the scalar field shares the Killing symmetry and since no peculiarities of the scalar field are pointed out, we assume that it is supposed to be regular, in particular at the event horizon. But in our case due to the constraint (\ref{cond_1}) the first of the assumptions may be violated, in addition the derivative of the scalar field has singular behaviour at the horizon, therefore the conclusions made in \cite{Sang_PRD21} cannot be applied directly to our solution. It was argued \cite{Hajian_PLB21} that in Horndeski theory instead of the standard surface gravity its ``effective" counterpart can be introduced and it can be explained by the fact that in general the speed of gravitons may differ form the speed of light \cite{Kobayashi_PRD14,Bettoni_PRD17}, namely these speeds differ if the Lagrangian for the gravitational perturbation contains Weyl tensor (the so-called Weyl criterion), what usually takes place in Horndeski case \cite{Bettoni_PRD17}. Consequently, the ``effective" or modified surface gravity gives rise to a modified relation for the black hole temperature \cite{Hajian_PLB21} which can be written in the form:
\begin{equation}\label{temp_gen}
T=\frac{\kappa}{2\pi}\left(1+\frac{\e}{4}\frac{(\vp')^2}{W}\right)\Big|_{r_{+}}=T_{BH}\left(1+\frac{\e}{4}\frac{(\vp')^2}{W}\right)\Big|_{r_{+}}=\sqrt{U(r_+)W(r_+)}T_{BH}.
\end{equation}
For the particular case of the solution given by the metric (\ref{metric}) with corresponding functions $U(r)$ and $W(r)$ we obtain:
\begin{equation}\label{temp}
T=\frac{\eta}{8\pi(n-1)\alpha r_{+}(r^2_{+}+d^2)}\left(\left(\frac{\al}{\eta}-\L\right)r^2_{+}+(n-1)(n-2)-\frac{q^2}{r^{2(n-2)}_{+}}-\frac{(n-1)(n-2)}{2r^2_{+}}\bar{q}^2\right)^2.
\end{equation}
We point out that in the limit $\eta\to 0$ the both the temperature (\ref{temp}) and its cousin (\ref{T_aux}) become singular, confirming the fact that our solution, from which both these expressions are derived, does not meet the criteria imposed in \cite{Sang_PRD21}. Even though the temperature (\ref{temp}) is given by a relatively simple expression not all its peculiarities can be seen easily, but nevertheless its key features can be described. First of all, due to the square over the main parentheses an effective coupling between the terms of different origin appear, namely there is a coupling between both gauge fields given by the term proportional to $q^2\bar{q}^2$, but we can also claim about a ``coupling" between the gauge and scalar fields reflected by the terms where coupling parameters are multiplied by $q^2$ or $\bar{q}^2$. To sum it up the coupling we mention here is just a consequence of the coupling caused by Horndeski gravity and which appear in the metric functions $U(r)$ and $W(r)$. 

Using the relation (\ref{temp}) we can easily analyse  asymptotic behaviour of the temperature. For instance for large $r_+$ ($r_+\to\infty$) the temperature $T$ (\ref{temp}) shows de Sitterian or anti-de Sitterian character depending on signs of the parameters $\al$ and $\e$, namely $T\sim (\al-\L\e)^2r_+/(2(n-1)\al\e)$,  but we pay more attention to the former one, de Sitterian case will be examined elsewhere. For very small $r_+$ ($r_+\to 0$) the temperature is mainly defined by the gauge field terms, and if $n>3$ the leading term is related to the Maxwell field and it is of the form as follows $T\sim q^4/((n-1)^2(n-2))r^{7-4n}_{+}$, what is curious here, being caused by the nonminimal coupling this leading term does not have any dependence neither of the parameter $\e$ nor of the parameter $\al$. If $n=3$ both gauge field terms give equal contribution, due to their symmetry even in the metric (\ref{U_n3}) and consequently it is reflected in temperature. 

\begin{figure}
\centerline{\includegraphics[scale=0.33,clip]{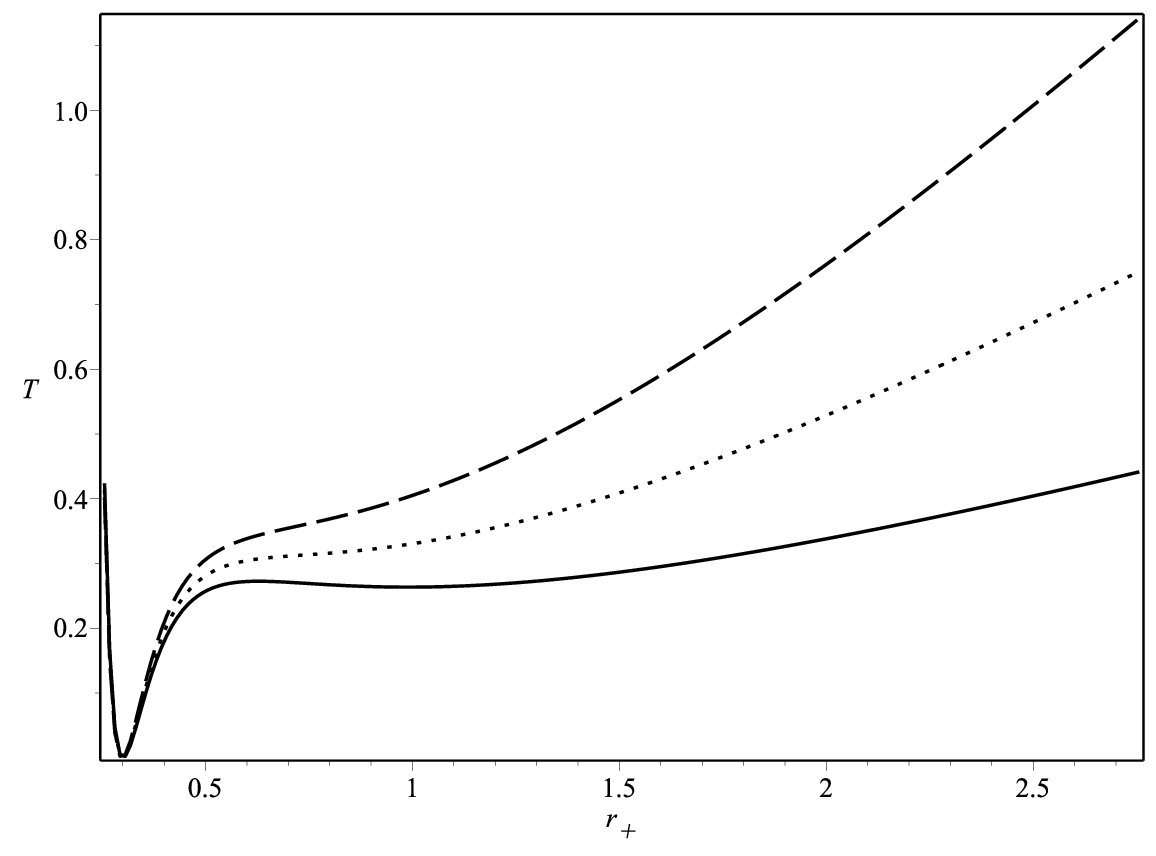}
\includegraphics[scale=0.36,clip]{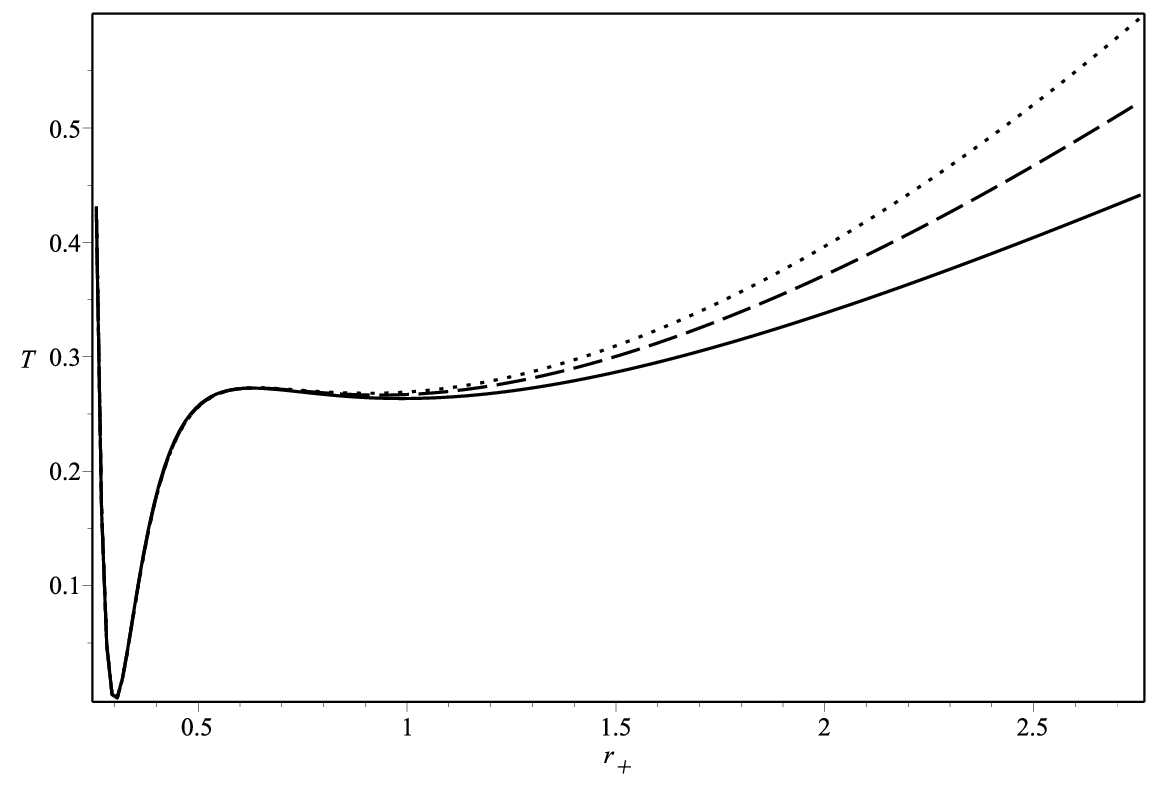}}
\caption{Black hole's temperature $T$ as a function of horizon radius  $r_+$ for some values of the cosmological constant $\L$ (the left graph) and the coupling parameter $\e$ (the right one). For both cases the way ``form bottom to top" corresponds to the increase in absolute value of the parameter we vary, whereas all other parameters are held fixed, namely for both graphs we have taken $n=4, \al=0.1, q=\bar{q}=0.2$. For the left graph we take $\e=0.2$ and $L_1=-2$, $\L_2=-3$ and $\L_3=-4$, whereas for the right graph we take $\L=-2$ and $\e_1=0.2$, $\e_2=0.4$ and $\e_3=0.8$.}\label{temp_fig_1}
\end{figure}

The analysis of the temperature as a function of the horizon radius for its intermediate values is not trivial since the contribution of various terms may be comparable what affects on the behaviour of the temperature. The terms in the relation (\ref{temp}) have opposite signs the temperature might be a nonmonotonous function of the horizon radius $r_+$. In order to understand the dependence $T=T(r_+)$ better, we give some plots of this function for various values of parameters. The Figure  [\ref{temp_fig_1}] shows this dependence if the cosmological constant $\L$ (the left graph) and the parameter of nonminimal coupling $\e$ (the right graph) are varied. The general features of both graphs are very similar, the function $T=T(r_+)$ has a specific ``narrow'' and ``deep'' minimum, this minimum is not affected considerably by variation of either $\L$ or $\e$ and we conclude that it is mainly defined by the gauge field terms (it is given below). If the cosmological constant rises in absolute value the temperature $T$ also rises for large $r_{+}$ and it tends to be more monotonous in the range of intermediate values of $r_{+}$. The mentioned feature is also known for Reissner-Nordstrom-AdS black hole and it causes nontrivial critical behaviour within extended thermodynamics approach, the latter will be considered in the following sections. Comparing both graphs of the Fig.[\ref{temp_fig_1}] we also conclude that variation of the cosmological constant $\L$ leads to more substantial change of the temperature for intermediate and relatively large values of the horizon radius $r_+$ than the variation of the coupling constant $\eta$ and this result is expectable, because of the way those parameters contribute into the expression (\ref{temp}).

\begin{figure}
\centerline{\includegraphics[scale=0.36,clip]{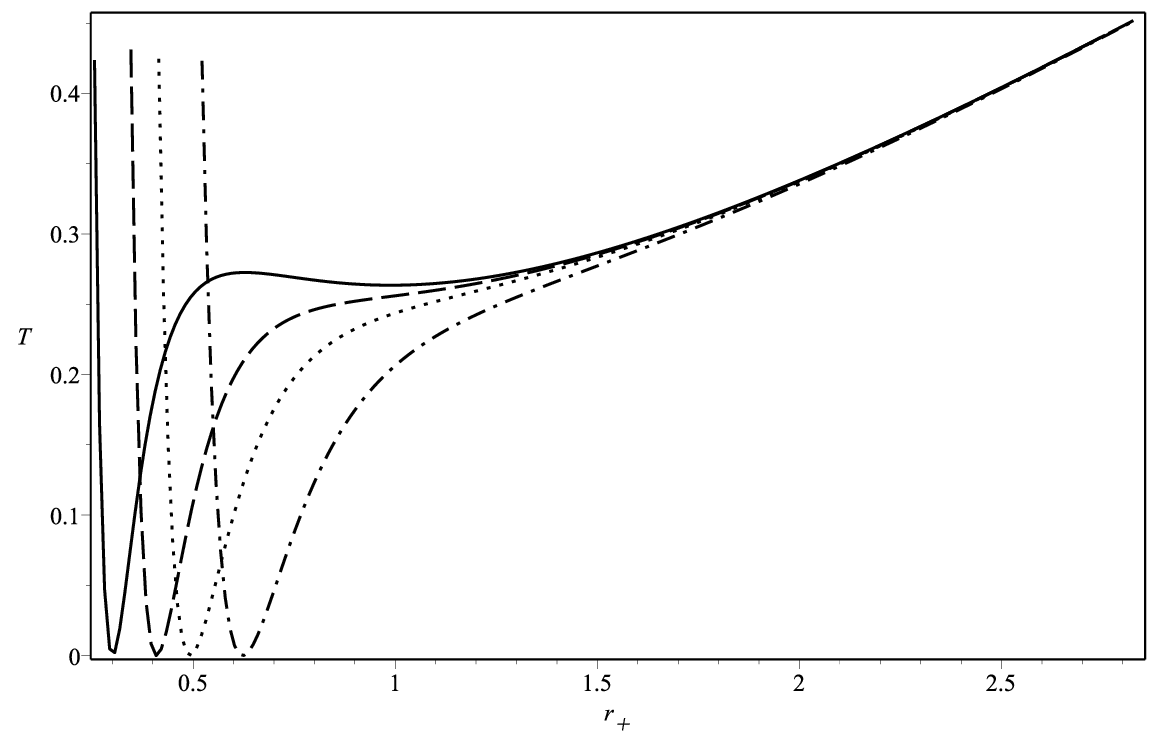}}
\caption{Black hole's temperature $T$ as a function of horizon radius  $r_+$ for some values of the electric charge $q$ if all the other parameters are held fixed. The solid, dashed, dotted and dash-dotted curves correspond to $q_1=0.2$, $q_2=0.4$, $q_3=0.6$ and $q_4=1$ respectively. The fixed parameters are as follows: $n=4$, $\al=0.1$, $\e=0.2$, $\L=-2$  and $\bar{q}=0.2$.}\label{temp_fig_2}
\end{figure}

The Figure [\ref{temp_fig_2}] shows the influence of variation of the electric charge $q$ on the temperature $T$. As it is pointed out above since the terms caused by the gauge field become principal ones for small radii of horizon $r_+$ and it gives rise to the shift of the global minimum to the right if the charge $q$ goes up. We also point out that the ``narrow''  domain close to the global minimum changes considerably, namely it widens if the charge $q$ increases. The other important consequence of this variation is the fact that domain right to the global minimum also changes substantially, namely its nonmonotonicity becomes less notable and we can conclude that further increase of the charge gives rise to its disappearance and it also affects on the critical behaviour of the black hole. Due to the same sign and inverse proportionality to the horizon radius $r_+$ a variation of the nonabelian charge $\bar{q}$ gives qualitatively to the same changes in behaviour of the temperature $T$, but due to different $r_+$ dependences in general case, those changes might be substantial for intermediate values of $r_+$. Just for particular case $n=3$ both gauge field give equal contribution.

\section{Wald procedure, conserved quantities and the first law of black hole thermodynamics}
Wald approach is a consistent method to derive the first law of black hole mechanics (thermodynamics). Being a generalization of the standard Noether procedure to obtain conserved quantities it allows to gain the latter one for general diffeomorphism invariant theories and it was successfully applied to various gravity theories, moreover the approach was generalized for the theories with internal gauge degrees of freedom \cite{Prabhu_CQG17}. To briefly describe the procedure we write the variation of the Lagrangian for the system (\ref{action}):
\begin{equation}\label{var_lagr}
\de{\cal L}=\sqrt{-g}\left({\cal E}_{\mu\nu}\de g^{\mu\nu}+{\cal E}_{\vp}\de\vp+{\rm Tr}\left({{\cal E}_{A}}^{(a)\mu}\de A^{(a)}_{\mu}\right)+{{\cal E}_{\cal A}}^{\mu}\de {\cal A}_{\mu}\right)+\sqrt{-g}\nb_{\mu}{\cal J}^{\mu},
\end{equation}
and here ${\cal E}_{\mu\nu}$, ${\cal E}_{\vp}$, ${{\cal E}_{A}}^{(a)\mu}$ and ${{\cal A}}^{\mu}$ are the left hand sides of the equations of motion (\ref{einstein}), (\ref{scal_f_eq}), (\ref{YM_eq}) and (\ref{em_eq})  respectively for the dynamical fields we consider. The last term in the upper variation is the so-called boundary term which is transformed into a hypersurface integral enclosing the chosen volume and ${\cal J}^{\mu}$ is the surface current which can be given as a sum of corresponding dynamical fields currents, namely:
\begin{equation}\label{surf_curr}
{\cal J}^{\mu}={\cal J}^{\mu}_{g}+{\cal J}^{\mu}_{\vp}+{\cal J}^{\mu}_{A}+{\cal J}^{\mu}_{{\cal A}},
\end{equation}
where the respective components are defined as follows:
\begin{gather}
{\cal J}^{\mu}_{g}=2\frac{\partial {\cal L}}{\partial R_{\k\l\mu\nu}}\nb_{\l}\left(\de g_{\k\nu}\right)-2\nb_{\l}\left(\frac{\partial {\cal L}}{\partial R_{\k\mu\l\nu}}\right)\de g_{\k\nu},\\{\cal J}^{\mu}_{\vp}=\frac{\partial {\cal L}}{\partial(\vp_{\mu})}\de\vp, \quad  {\cal J}^{\mu}_{A}=-4{\rm Tr}\left(F^{(a)\mu\l}\de A^{(a)}_{\l}\right), \quad {\cal J}^{\mu}_{{\cal A}}=-4{\cal F}^{\mu\l}\de{\cal A}_{\l}.
\end{gather}
If the equations of motion are satisfied the only contribution to the variation of the Lagrangian (\ref{var_lagr}) and respectively to the action is given by the hypersurface term. Having the current ${\cal J}^{\mu}$ (\ref{surf_curr}) we can construct corresponding current form $J_{(1)}={\cal J}_{\mu}dx^{\mu}=g_{\l\mu}{\cal J}^{\l}dx^{\mu}$ and then we define its Hodge dual which is essential in Wald approach:
\begin{equation}\label{hodge_curr}
\Theta(\psi,\de\psi)=*J_{(1)}(\psi,\de\psi),
\end{equation}
where $\psi$ is used to denote all the dynamical fields and $\de\psi$ is their variations. The diffeomorphism is generated by a vector field $\xi^{\mu}$, therefore the variation of dynamical fields can be written in the form:
\begin{equation}\label{var_lie}
\de_{\xi}\psi={\cal L}_{\xi}\psi,
\end{equation}
where ${\cal L}_{\xi}$ is the corresponding Lie derivative, generated by the vector $\xi^{\mu}$. The variation of the Lagrangian of the system can be written also as corresponding Lie derivative, namely:
\begin{equation}\label{var_lagr_form}
\de_{\xi}{\cal *L}={\cal L}_{\xi}{\cal *L}=d(i_{\xi}*{\cal L}),
\end{equation}
here we point out that since the Lagrangian in our case is defined as a scalar function {\it i.e.} $0$-form, therefore in the latter relation the Hodge dual of the Lagrangian is used and we note that to derive the second equality the upper relation the so-called Cartan magic formula is used. Rewriting the formula for the variation of the Lagrangian (\ref{var_lagr}) in terms of forms and taking into account the relations (\ref{hodge_curr}) (\ref{var_lagr_form}) as well as the notation (\ref{var_lie}) we obtain:
\begin{equation}
d(i_{\xi}{\cal *L})={\cal E}_{\psi}{\cal L}_{\xi}\psi+d\Theta(\psi,{\cal L}_{\xi}\psi) \quad \Rightarrow \quad d\left(\Theta(\psi,{\cal L}_{\xi}\psi)-i_{\xi}{\cal *L}\right)=-{\cal E}_{\psi}{\cal L}_{\xi}\psi,
\end{equation}
where ${\cal E}_{\psi}$ correspond to the equations of motions for the dynamical fields. If the equations of motion are satisfied, the right hand side of the latter reation will be equal to zero. Now we introduce a Noether current $n$-form:
\begin{equation}\label{N_current}
J_{\xi}=\Theta(\psi,\de\psi)-i_{\xi}{\cal *L},
\end{equation}
which is obviously closed on-shell, moreover it implies that this form is exact on-shell, namely:
\begin{equation}\label{N_charge}
J_{\xi}=dQ_{\xi}.
\end{equation}
Integral over a closed $n-1$ dimensional hypersurface $\Sigma_{n-1}$ is the so-called Noether charge related to the vector field $\xi^{\mu}$ which generates the diffeomorphism. Then following Wald approach the space of the solutions of the equations of motion is defined to be the phase space of the theory and variation of the dynamical fields $\de_{\xi}\psi$ taken on-shell is a phase space vector flow generated by the vector $\xi^{\mu}$. This flow can be generated by a Hamiltonian ${\cal H}_{\xi}$ which is related to a symplectic form defined on a Cauchy hypersurface $\Sigma$, namely for its on-shell variation we write:
\begin{equation}
\de{\cal H}_{\xi}=\int_{\Sigma}\Omega(\psi,\de\psi,{\cal L}_{\xi}\psi)=\int_{\Sigma}\left(\de\Theta(\psi,{\cal L}_{\xi}\psi)-{\cal L}_{\xi}\Theta(\psi,\de\psi)\right).
\end{equation}
Using the definition of the Noether current (\ref{N_current}) and Cartan magic formula for the Lie derivative we can rewrite the latter relation as follows:
\begin{equation}
\de{\cal H}_{\xi}=\int_{\Sigma}\left(\de J_{\xi}+\de(i_{\xi}{\cal *L})-i_{\xi}d\Theta-d(i_{\xi}\Theta)\right)=\int_{\Sigma}(\de(dQ_{\xi})-d(i_{\xi}\Theta))=\int_{\partial\Sigma}\left(\de Q_{\xi}-i_{\xi}\Theta\right).
\end{equation}
We note that in the second equality we have used the on-shell condition which allow to remove the second and the third terms in the first integral. In the second integral we use the definition of the Noether charge and the fact that exterior derivative and the variation for the Noether charge $Q_{\xi}$ commute allowed to derive the last equality and the integral over the boundary $\partial\Sigma$. If $\xi^{\mu}$ is supposed to be a generator of a symmetry then ${\cal L}_{\xi}\phi=0$ and consequently $\de{\cal H}_{\xi}=0$. If the hypersurface $\Sigma$ has two boundaries, what actually takes place for black holes, namley the infinity and the event horizon, therefore from upper relation we obtain:
\begin{equation}\label{var_diff}
\de{\cal H}_{r_+}\equiv\int_{\partial\Sigma_+}\left(\de Q_{\xi}-i_{\xi}\Theta\right)=\int_{\infty}\left(\de Q_{\xi}-i_{\xi}\Theta\right)\equiv\de{\cal H}_{\infty}
\end{equation}
and here $\partial\Sigma_{+}$ is the event horizon hypersurface. The written relation allows to derive the first law of black hole thermodynamics.

Before derivation of the first law of black hole thermodynamics we give an explicit relation for the components of the Noether charge, namely we write:
\begin{equation}
Q_{\l_1\ldots\l_{n-1}}=\ve_{\l_1\ldots\l_{n-1}\mu\nu}\left(\frac{\partial {\cal L}}{\partial R_{\k\l\mu\nu}}\nb_{\l}\xi_{\k}-2\xi_{\k}\nb_{\l}\left(\frac{\partial {\cal L}}{\partial R_{\k\l\mu\nu}}\right)-2{\rm Tr}\left(F^{(a)\mu\nu}A^{(a)}_{\l}\right)\xi^{\l}-2{\cal F}^{\mu\nu}{\cal A}_{\l}\xi^{\l}\right).
\end{equation} 
Using the upper relation as well as the relation for the Hodge dual of the surface current (\ref{hodge_curr}) we can calculate the differences of variations which are given under the integrals in the relation (\ref{var_diff}). Similarly as in the previous section the time translation vector $\xi^{\mu}$ can be chosen for corresponding calculations, it is a Killing vector and it is null on the event horizon. For more clarity we split the calculations of the difference of the variations on two parts, namely for the gravity part together with nonminimally coupled scalar field and for the gauge fields.  The gravity part together with the scalar field contribution give rise to the following relation:
\begin{equation}\label{var_gr}
\left(\de Q_{\xi}-i_{\xi}\Theta\right)_{gs}=-(n-1)r^{n-2}\de U \hat{\Omega}_{n-1},
\end{equation}
where $\de U$ is the variation of the metric function $U$ and $\hat{\Omega}_{n-1}$ is the surface $n-1$-form. The total variation for nonminimally coupled theory excluding gauge field contribution depends on the variation of the metric function $\de U$ only and we point out that similar result is derived in pure Einsteinian theory for instance for the Schwarzschild solution. The gauge fields give independent contribution and it takes the form:
\begin{equation}\label{var_gauge}
\left(\de Q_{\xi}-i_{\xi}\Theta\right)_{gf}=\frac{2r^{n-1}}{\sqrt{UW}}
{\cal A}_{0}\left(\left(\frac{\de U}{U}+\frac{\de W}{W}\right){\cal A}'_{0}-2\de{\cal A}'_{0}\right)\hat{\Omega}_{n-1},
\end{equation}
where ${\cal A}_{0}$ is the time component of the electromagnetic field potential and ${\cal A}'_{0}={\cal F}_{rt}$ is its radial derivative (electric field). We would like to stress that the Yang-Mills field does not give any contribution to the difference of variations due to the fact that the constant $\bar{q}$ associated with Yang-Mills coupling is held fixed. The total variation is the sum of the both written above variations:
\begin{equation}
\left(\de Q_{\xi}-i_{\xi}\Theta\right)_{tot}=r^{n-2}\left(-(n-1)\de U+\frac{2r}{\sqrt{UW}}
{\cal A}_{0}\left(\left(\frac{\de U}{U}+\frac{\de W}{W}\right){\cal A}'_{0}-2\de{\cal A}'_{0}\right)\right)\hat{\Omega}_{n-1}.
\end{equation} 
For convenience we assume that the electric potential is equal to zero at the event horizon ${\cal A}_{0}|_{r_{+}}=0$. Taking this condition into account and performing integration over a $n-1$-dimensional hypersphere of the radius $r_+$ we obtain the explicit relation for the variation of the Hamiltonian ${\cal H}_{r_+}$ at the horizon: 
\begin{equation}
\de{\cal H}_{r_+}=(n-1)\omega_{n-1}r^{n-2}_{+}U'(r_+)\de r_{+},
\end{equation}
where $\omega_{n-1}=2\pi^{(n-1)/2}/\Gamma((n-1)/2)$ is the surface are of a unit $n-1$ dimensional hypersphere. Variation of the Hamiltonian $\de{\cal H}_{\infty}$ takes the form as follows:
\begin{equation}
\de{\cal H}_{\infty}=(n-1)\omega_{n-1}\de\mu-4\omega_{n-1}{\cal A}_{0}\de q.
\end{equation}
Since as it is pointed out above the variation of the Hamiltonian at the horizon and at the infinity are equal, therefore we obtain:
\begin{equation}\label{FL_bare}
(n-1)\omega_{n-1}r^{n-2}_{+}U'(r_+)\de r_{+}=(n-1)\omega_{n-1}\de\mu-4\omega_{n-1}{\cal A}_{0}\de q.
\end{equation} 
Finally, to derive the first law of black hole thermodynamics, it is necessary to find the relations between the variations of observable entities such us mass or charge of the black hole and corresponding variations in the given above relation.

The electric charge is defined in the standard way, namely we use the Gauss law and obtain:
\begin{equation}\label{el_charge}
Q_{e}=\frac{1}{4\pi}\int_{\Sigma_{\infty}}*F=\frac{\omega_{n-1}}{4\pi}q.
\end{equation}
The electric potential measured at the infinity with respect to the horizon is defined as follows:
\begin{equation}\label{el_pot}
\Phi_{e}={\cal A}_{\mu}\xi^{\mu}|_{\infty}-{\cal A}_{\mu}\xi^{\mu}|_{r_{+}}={\cal A}_{0}.
\end{equation}  
we point out that the time translation vector $\xi^{\mu}=\partial/\partial t$ is used here to calculate the electric potential. Black hole's mass can be defined as:
\begin{equation}\label{mass}
M=\frac{(n-1)\omega_{n-1}}{16\pi}m.
\end{equation}
Variation of the mass (\ref{mass}) together with the relations (\ref{el_charge}) and (\ref{el_pot}) allow to rewrite the right hand side of the equation (\ref{FL_bare}) in a form of a typical thermodynamic relation.  In the left hand side of that relation we can use the relation for the temperature (\ref{temp}) in order to avoid introducing additional scalar charges and its corresponding conjugate value if the physical meaning of both these values is not clarified. Then the entropy of the black hole can be defined in a typical manner, namely:
\begin{equation}\label{entropy}
S=\frac{\omega_{n-1}}{4}r^{n-1}_{+}.
\end{equation}
Therefore, the entropy is equal to a quarter of the black hole horizon area, similarly as it takes place in General Relativity. Finally, the first law of the black hole thermodynamics can be written in the form:
\begin{equation}\label{FL}
\delta M=T\delta S+\Phi_{e}\delta Q_{e}.
\end{equation}
The obtained relation is completely of the same form as for the Reissner-Nordstrom black hole in the framework of GR, even though the explicit relation for the temperature (\ref{temp}) differs from its general relativistic cousin. The fact that the thermodynamic relations like the first law are the same in different theories may be an additional confirmation of universality of black hole thermodynamics which at least for some cases are insensitive to the underlying theories which allow to obtain corresponding thermodynamic relations. 

We would also like to stress that even from a naive thermodynamic point of view the temperature $T$ (\ref{temp}) satisfies a simple consistency relation, which follows directly from the first law (\ref{FL}), namely $\frac{\partial T}{\partial Q_{e}}=\frac{\partial \Phi_{e}}{\partial S}$ whereas the temperature $T$ (\ref{T_aux}) does not. To obtain consistency relation for the temperature (\ref{T_aux}) additional scalar charge was introduced \cite{Feng_PRD16}, what was used in earlier paper \cite{Stetsko_PRD19,Stetsko_PRD20}, but its physical meaning is not clear.  Moreover in the framework of the standard thermodynamics there are only two variable macroscopic parameters of the black hole, namely its mass or directly related to it the radius of the event horizon $r_+$ and the electric charge $q$ (or $Q_{e}$), any additional independent thermodynamic variable should be related to an independent macroscopic parameter (integration constant), but there are no any more independent macroscopic values in the standard framework. Thus, the ``scalar charge" considered in earlier paper was introduced just to have consistent thermodynamics relations, but its physical meaning remains obscure. 

Heat capacity or specific heat is an important notion to analyze thermal stability, particularly it is widely used in black hole thermodynamics. Thermally stable systems are characterized by positive specific heat and if the specific heat turns to be negative the system tends to decay. To obtain the heat capacity we use the standard definition for the latter and write:
\begin{multline}\label{heat_capac}
C_{Q}=T\left(\frac{\partial S}{\partial T}\right)_{Q}=T\frac{\partial S}{\partial r_{+}}\left(\frac{\partial r_{+}}{\partial T}\right)_{Q}=\frac{(n-1)\omega_{n-1}}{4}r^{n-2}_{+}\left(\left(\frac{\alpha}{\eta}-\Lambda\right)r^2_{+}+(n-1)(n-2)-\right.\\\left.\frac{q^2}{r^{2(n-2)}_{+}}-\frac{(n-1)(n-2)\bar{q}^2}{2r^2_{+}}\right)\left[-\frac{3r^2_{+}+d^2}{r_{+}(r^2_{+}+d^2)}\left(\left(\frac{\al}{\eta}-\Lambda\right)r^2_{+}+(n-1)(n-2)-\right.\right.\\\left.\left.\frac{q^2}{r^{2(n-2)}_{+}}-\frac{(n-1)(n-2)\bar{q}^2}{2r^2_{+}}\right)+4\left(\left(\frac{\al}{\eta}-\Lambda\right)r_{+}+\frac{(n-2)q^2}{r^{2n-3}_{+}}+\frac{(n-1)(n-2)\bar{q}^2}{r^3_{+}}\right)\right]^{-1}.
\end{multline}
The obtained relation (\ref{heat_capac}) has a relatively more cumbersome structure in comparison with the expression for the temperature (\ref{temp}), but since the derivative of the temperature $T$ over the horizon radius $r_{+}$ makes contribution in the heat capacity some important conclusions about the behaviour of the latter can be derived immediately knowing the peculiar features of the temperature. Namely, since the temperature may in general have three extrema points it means that the heat capacity as a function of $r_{+}$ may have three discontinuity points, separating stable and unstable domains. We point out here that since for relatively large $r_+$ the temperature shows rising character for any variation of black hole parameters, at least in the observed domain, therefore we can conclude that the specific heat $C_Q$ is positive and the black hole is thermally stable. For smaller radii of the horizon the sign of $C_Q$ and consequently conclusion about thermal stability or instability substantially depend of the chosen values of black hole parameters and the parameters of the Lagrangian. To make the behaviour of the function $C_Q=C_Q(r_{+})$ more transparent we give corresponding graphs, showing its behaviour near discontinuity points and how it is affected by variations of certain parameters namely its electric charge $q$ and the cosmological constant $\Lambda$. 

\begin{figure}
\centerline{\includegraphics[scale=0.38,clip]{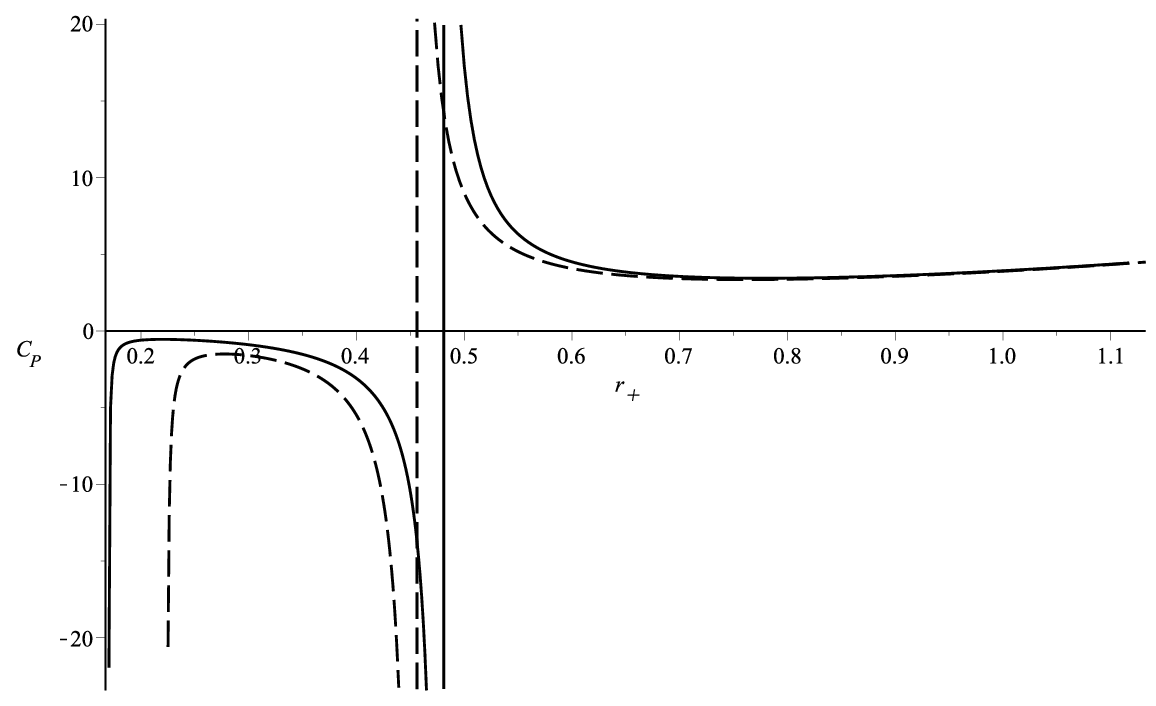}}
\caption{Heat capacity $C_Q$ as a function of horizon radius  $r_+$ for two values of the electric charge $q$ whenever the other parameters are held fixed. The solid and dashed curves correspond to $q=0.1$, and $q=0.125$ respectively. The fixed parameters are as follows: $n=3$, $\al=0.2$, $\e=0.3$, $\L=-3$  and $\bar{q}=0.01$.}\label{Cp_fig}
\end{figure}

\begin{figure}
\centerline{\includegraphics[scale=0.38,clip]{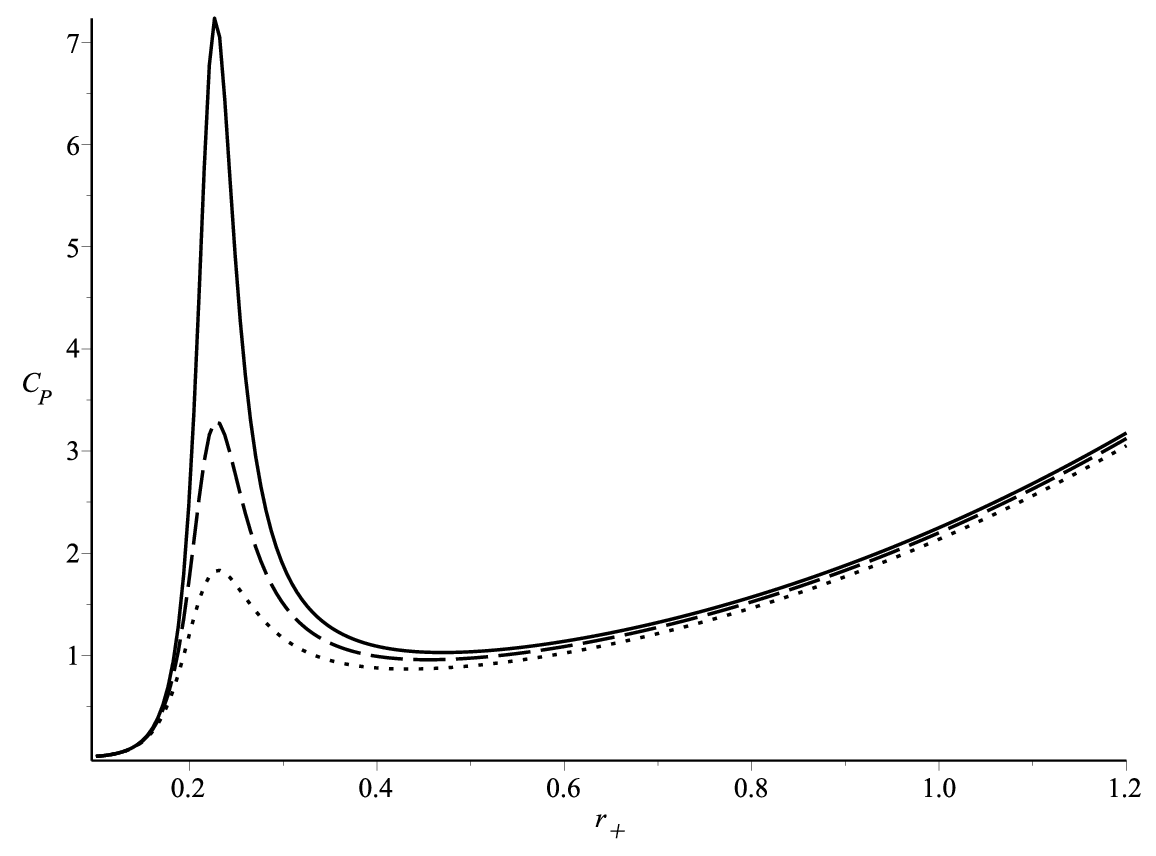}}
\caption{Dissolution of the peak for the heat capacity $C_{Q}$ for large absolute values of $\L$. The peaks ``higher to lower" correspond to increase of the absolute value of the cosmological constant.}\label{Cp_L_fig}
\end{figure}

The Figure~[\ref{Cp_fig}] shows the rightmost discontinuity point for two values of the electric charge. As it was noted above the heat capacity $C_{Q}$ to the right of the discontinuity point is positive and it goes up if the horizon radius $r_+$ increases, this feature is typical for most types of black holes with AdS asymptotic. Left to the asymptotes the heat capacity becomes negative, therefore this range of $r_+$ is a domain of instability. We also point out that for smaller radius $r_+$ there is a second discontinuity point what is reflected by very fast decrease of the heat capacity $C_{Q}$ if the radius of the horizon goes down. We also conclude that discontinuity points become closer if the charge $q$ goes up, and further increase of the charge gives rise to merging of the singularity points and consecutive shrinkage of the unstable domain at least for the considered range of the parameters. Similar conclusion can be made if the absolute value of $\L$ goes up. Then the peculiarity of the heat capacity diminishes what is shown on the Figure~[\ref{Cp_L_fig}], namely the height of the peak drops down and finally it vanishes if the absolute value of the cosmological constant $\L$ rises. 

We also point out that the heat capacity $C_Q$ (\ref{heat_capac}) within the extended thermodynamics approach can be treated  as the heat capacity under constant pressure $C_P$, and the pressure is introduced below. It is valid since all the parameters are held fixed in the relation (\ref{heat_capac}).

\section{Extended thermodynamics} 
The so-called extended thermodynamics attracts considerable attention during for more than a decade \cite{Kubiznak_JHEP12,Gunasekaran_JHEP12,Kubiznak_CQG2017}. Even though some basic assumptions for the extended thermodynamics are still disputed, this approach gives rise to wider thermodynamic phase space allowing to describe richer thermodynamics and establish at least formal, but deeper ties with the thermodynamics of various systems usually considered in condensed matter physics. In particular, it establishes profound relations between phase transition phenomena of condensed matter systems and phase transitions (transformations) in black hole physics. The key assumption of the extended approach is the fact that the cosmological constant is considered to be a thermodynamic value.  Namely, the cosmological constant $\Lambda$ was identified with thermodynamic pressure:
\begin{equation}\label{press}
P=-\frac{\Lambda}{8\pi}
\end{equation}
It should be pointed out that there is some analogy with ideal fluid, where a corresponding term related to the thermodynamic pressure goes along with the metric tensor in energy-momentum tensor of the fluid, but it will not be discussed in the current work. The introduced thermodynamic pressure (\ref{press}) gives rise to the consequence that the black hole mass should be identified now with the enthalpy $M=H$ \cite{Kubiznak_JHEP12}, but not with the internal energy as it was in the standard thermodynamics. Having the pressure $P$ (\ref{press}) the corresponding conjugate thermodynamic volume $V$ can be defined as follows:
\begin{equation}\label{TD_vol}
V=\left(\frac{\partial M}{\partial P}\right)_{S,Q_{e}}
\end{equation}
The explicit relation for the thermodynamic volume depends on the parity of dimension $n$, similarly as it is for the metric function $U(r)$ (\ref{U_mf}). Namely, for odd $n$ the explicit expression for the thermodynamic volume $V$ is as follows:
\begin{multline}\label{V_expl_odd}
V=\om_{n-1}\left(\frac{r^n_{+}}{n}-\frac{\eta}{4\al}\left(2\left(\frac{\alpha}{\eta}+\L\right)\left[\sum^{\frac{n-1}{2}}_{j=0}(-1)^jd^{2j}\frac{r^{n-2j}_{+}}{n-2j}+(-1)^{\frac{n+1}{2}}d^n\arctan{\left(\frac{r_+}{d}\right)}\right]+2q^2\times\right.\right.\\\left.\left.\left[\sum^{\frac{n-5}{2}}_{j=0}(-1)^jd^{2j}\frac{r^{4+2j-n}_{+}}{(4+2j-n)d^{2(j+1)}}+\frac{(-1)^{\frac{n-3}{2}}}{d^{n-2}}\arctan{\left(\frac{r_+}{d}\right)}\right]+(n-1)(n-2)\bar{q}^2\times\right.\right.\\\left.\left.\left[\sum^{\frac{n-5}{2}}_{j=0}(-1)^jd^{2j}\frac{r^{n-2j-4}_{+}}{n-2j-4}+(-1)^{\frac{n-3}{2}}d^{n-4}\arctan{\left(\frac{r_+}{d}\right)}\right]\right)\right).
\end{multline}
Explicit expression for the thermodynamic volume can be written similarly for even $n$. The obtained relation (\ref{V_expl_odd}) is in agreement with a respective relation obtained in \cite{Stetsko_PRD19} for corresponding limits in both cases.  Since $n=3$ is of special interest we also write the termodynamic volume for this case:
\begin{equation}\label{vol_n3}
V=4\pi\left(\frac{r^3_{+}}{3}-\frac{\eta}{2\al}\left[\left(\frac{\al}{\eta}+\L\right)\left(\frac{1}{3}r^3_{+}-d^2r_{+}\right)+\left(d^3\left(\frac{\al}{\eta}+\L\right)+\frac{1}{d}(q^2+\bar{q}^2)\right)\arctan{\left(\frac{r_+}{d}\right)}\right]\right).
\end{equation}
To derive the Smarr relation for the black hole we introduce additional intensive thermodynamic variable, which in some sense is similar to the pressure (\ref{press}) introduced above. The new variable and its conjugate are defined as follows:
\begin{equation}\label{add_var}
\Pi=\frac{\al}{8\pi\eta}, \quad \Psi=\left(\frac{\partial M}{\partial \Pi}\right)_{S,Q_{e},P} 
\end{equation}
Taking correponding derivatives we write the explicit relation for the extensive conjugate value $\Psi$. Namely, for odd $n$ ($n<7$) we obtain:
\begin{multline}\label{Psi_n}
\Psi=\om_{n-1}\left(\frac{\eta}{2\al}\left[\left(\frac{\al}{\eta}+L\right)\left(\sum^{\frac{n-1}{2}}_{j=0}(-1)^jd^{2j}\frac{r^{n-2j}_{+}}{n-2j}+(-1)^{\frac{n-1}{2}}d^n\arctan{\left(\frac{r_+}{d}\right)}\right)+(n-1)(n-2)\frac{\bar{q}^2}{2}\times\right.\right.\\\left.\left.\left(\sum^{\frac{n-5}{2}}_{j=0}(-1)^jd^{2j}\frac{r^{n-2j-4}_{+}}{n-2j-4}+(-1)^{\frac{n-3}{2}}d^{n-4}\arctan{\left(\frac{r_+}{d}\right)}\right)+q^2\left(\sum^{\frac{n-5}{2}}_{j=0}(-1)^j\frac{r^{4+2j-n}_{+}}{(4+2j-n)d^{2(j+1)}}+\right.\right.\right.\\\left.\left.\left.\frac{(-1)^{\frac{n-3}{2}}}{d^{n-2}}\arctan{\left(\frac{r_+}{d}\right)}\right)\right]+\frac{\eta^2r^{n-2}_{+}}{4\al^2(r^2_{+}+d^2)}\left(\left(\frac{\al}{\eta}+\L\right)r^2_{+}+\frac{q^2}{r^{2(n-2)}_{+}}+\frac{(n-1)(n-2)}{2}\frac{\bar{q}^2}{r^2_{+}}\right)^2-\frac{\eta^2}{8\al^2}\times\right.\\\left.\left[(n+1)\left(\frac{\al}{\eta}+\L\right)^2\left(\sum^{\frac{n-1}{2}}_{j=0}(-1)^jd^{2j}\frac{r^{n-2j}_{+}}{n-2j}+(-1)^{\frac{n-1}{2}}d^n\arctan{\left(\frac{r_+}{d}\right)}\right)+2(4-n)\left(\frac{\al}{\eta}+\L\right)q^2\times\right.\right.\\\left.\left.\left(\sum^{\frac{n-5}{2}}_{j=0}(-1)^j\frac{r^{4+2j-n}_{+}}{(4+2j-n)d^{2(j+1)}}+\frac{(-1)^{\frac{n-3}{2}}}{d^{n-2}}\arctan{\left(\frac{r_+}{d}\right)}\right)+3(2-n)q^4\left(\sum^{\frac{3n-7}{2}}_{j=0}(-1)^j\frac{r^{6+2j-3n}_{+}}{(6+2j-3n)d^{2(j+1)}}+\right.\right.\right.\\\left.\left.\left.\frac{(-1)^{\frac{3n-5}{2}}}{d^{3n-4}}\arctan{\left(\frac{r_+}{d}\right)}\right)+(n-1)(n-2)\left(\frac{\al}{\eta}+\L\right)\bar{q}^2\left(\sum^{\frac{n-5}{2}}_{j=0}(-1)^jd^{2j}\frac{r^{n-2j-4}_{+}}{n-2j-4}+(-1)^{\frac{n-3}{2}}d^{n-4}\right.\right.\right.\\\left.\left.\left.\times\arctan{\left(\frac{r_+}{d}\right)}\right)-n(n-1)(n-2)q^2\bar{q}^2\left(\sum^{\frac{n-1}{2}}_{j=0}(-1)^j\frac{r^{2j-n}_{+}}{(2j-n)d^{2(j+1)}}+\frac{(-1)^{\frac{n+1}{2}}}{d^{n+2}}\arctan{\left(\frac{r_+}{d}\right)}\right)\right.\right.\\\left.\left.+\frac{1}{4}(n-1)^2(n-2)^2(n-6)\bar{q}^4\left(\sum^{\frac{5-n}{2}}_{j=0}(-1)^j\frac{r^{n+2j-6}_{+}}{(n+2j-6)d^{2(j+1)}}+\frac{(-1)^{\frac{7-n}{2}}}{d^{8-n}}\arctan{\left(\frac{r_+}{d}\right)}\right)\right]\right).
\end{multline}
For the dimensions $n\geqslant 7$ the there is a different contribution in the bottom line of the above relation, it follows from the corresponding term in the metric function $U(r)$. The explicit expression for $\Psi$ if $n$ is even can be derived similarly. We also write the thermodynamic function $\Psi$ for $n=3$ case:
\begin{multline}\label{Psi_3}
\Psi=4\pi\left(\frac{\eta}{4\al d}\left(1-\frac{\eta\L}{\al}\right)(q^2+\bar{q}^2)\arctan{\left(\frac{r_+}{d}\right)}-\frac{\eta\L}{\al}\left(1-\frac{\eta\L}{\al}\right)\left(\frac{r^3_+}{3}-d^2r_{+}+d^3\arctan{\left(\frac{r_+}{d}\right)}\right)-\right.\\\left.\frac{3\eta^2}{8\al^2d^2}(q^2+\bar{q}^2)^2\left(\frac{1}{3r^3_+}-\frac{1}{d^2r_+}+\frac{1}{d^3}\arctan{\left(\frac{r_+}{d}\right)}\right)+\frac{\eta^2 r_{+}}{8\al^2(r^2_{+}+d^2)}\left(\left(\frac{\al}{\eta}+\L\right)r^2_{+}+\frac{q^2+\bar{q}^2}{r^2_{+}}\right)^2\right).
\end{multline}
Since nonabelian field is also included into the action, moreover it gives a contribution into the metric function $U(r)$ and all the derived quantities we assume that the nonabelian parameter $\bar{q}$ can be varied as well. We introduce nonabelian charge similarly as it was defined for instance in \cite{Stetsko_PRD20_2}:
\begin{equation}\label{YM_charge}
Q_{n}=\frac{1}{4\pi\sqrt{(n-1)(n-2)}}\int_{\Sigma_{n-1}}d^{n-1}\chi J(\chi)\sqrt{Tr{(F^{(a)}_{\mu\nu}F^{(a)\mu\nu})}}=\frac{\om_{n-1}}{4\pi}\bar{q}.
\end{equation}
The integral in upper relation is taken over a sphere enclosing the black hole and $J(\chi)$ denotes the Jacobian for the chosen spherical coordinates. The Yang-Mills charge $Q_{n}$ (magnetic) now can considered as a thermydynamic value similarly to the electric charge of the Maxwell field. Therefore thermodynamic conjugate value to the charge $Q_{n}$ can be introduced:
\begin{equation}\label{NA_pot}
U=\left(\frac{\partial M}{\partial Q_{n}}\right)_{S,Q_{e},P,\Pi}.
\end{equation}
We do not give explicit expression for the potential $U$, but it can be obtained easily. Having introduced additional thermodynamic variables such as $P$, $\Pi$, $Q_n$ and their thermodynamic conjugates we are able to write the so-called extended first law, which takes the form:
\begin{equation}\label{ext_FL}
\de M=T_{BH}\de S+\Phi_{e}\de Q_{e}+V\de P+\Psi\de \Pi+U\de Q_{n}.
\end{equation}
Taking into account the pairs of conjugate variables we also write the Smarr relation:
\begin{equation}\label{smarr}
(n-2)M=(n-1)T_{BH}S-2VP-2\Psi\Pi+(2-n)\Phi_{e}Q_{e}+UQ_{n}.
\end{equation}
If nonabelian field is set to zero ($\bar{q}=0$) the obtained relation is reduced to the corresponding equation derived for the electrically charged black hole in Horndeski gravity \cite{Stetsko_PRD19}.  If we compare with general relativistic case the Smarr relation (\ref{smarr}) and the generalized first law (\ref{ext_FL}) gain only one additional term caused by the thermodynamic variable $\Pi$ and its conjugate value $\Psi$. The latter two relations may be considered as an additional argument in favour of universality of black hole thermodynamics, which allows us to write the fundamental thermodynamic relations that take the same or at least very similar form for various underlying theories of gravity.  

\section{Gibbs free energy}
\begin{figure}
\centerline{\includegraphics[scale=0.33,clip]{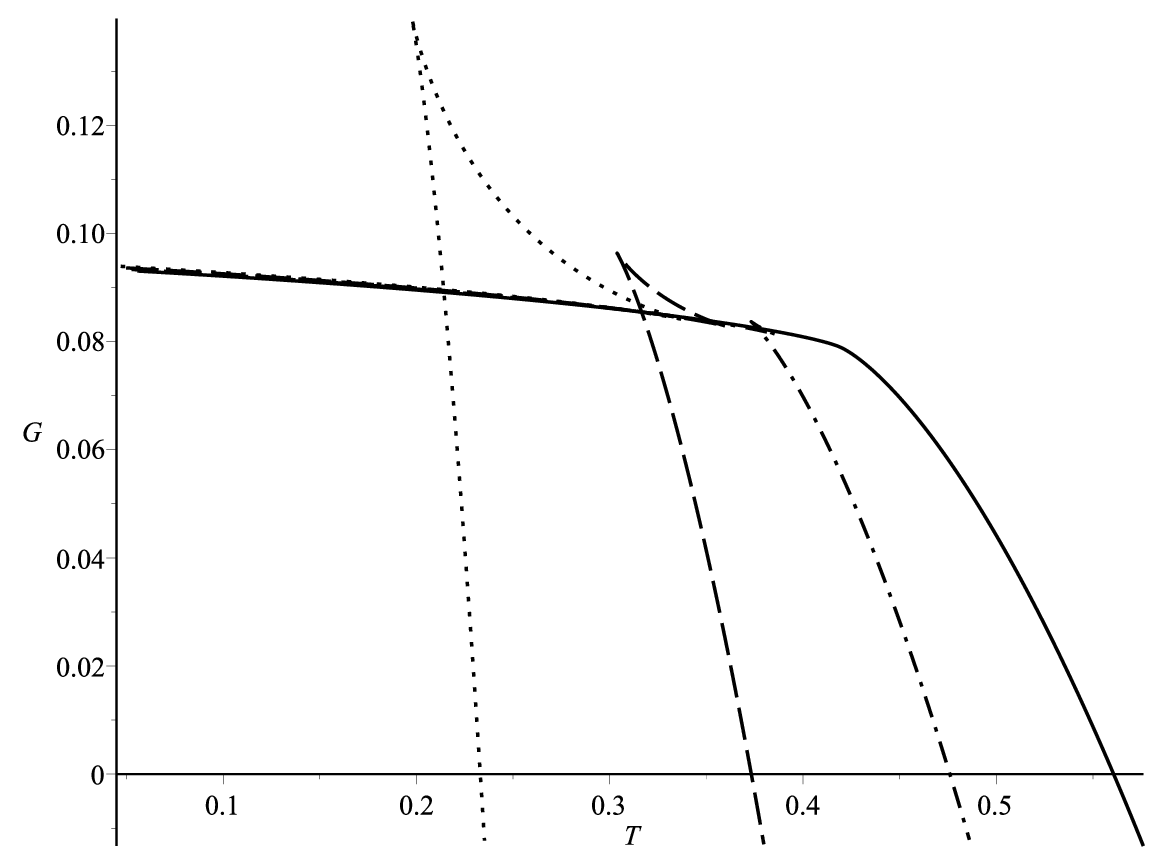}}
\caption{Gibbs free energy $G$ as a function of temperature $T$ for various values of pressure $P$ or cosmological constant $\L$, while other parameters are held fixed. The dotted, dashed, dash-dotted and solid lines correspond to $\L=-1.5$, $\L=-3.5$, $\L=-5.5$ and $\L=-7.5$ respectively. The fixed parameters are as follows: $n=3$, $q=0.1$, $\bar{q}=0.1$, $\al=0.2$ and $\e=0.3$.}\label{Gibbs_2d}
\end{figure}

\begin{figure}
\centerline{\includegraphics[scale=0.5,clip]{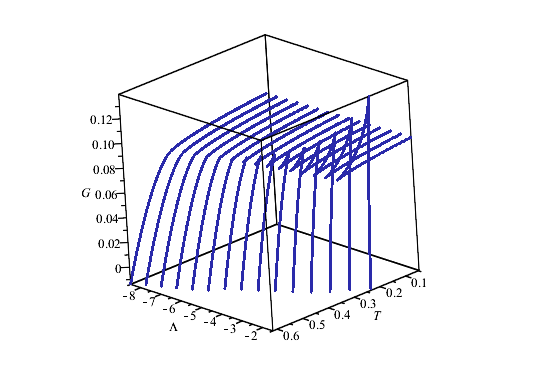}}
\caption{Gibbs free energy $G$ as a function of temperature $T$ and pressure $P$ (or the cosmological constant $\L$).}\label{Gibbs_3d}
\end{figure}

If a thermodynamic systems undergoes phase transitions the Gibbs free energy is much convenient than the enthalpy identified with the black hole's mass $M$. The Gibbs free energy is defined as follows:
\begin{equation}\label{Gibbs_def}
G=M-T_{BH}S.
\end{equation}
The explicit relation for the Gibbs free energy for odd $n$ ($n<7$) takes the forrm:
\begin{multline}\label{Gibbs_nod}
G=\frac{\om_{n-1}}{16\pi}\left(r^{n-2}_{+}+\frac{2\L}{n(n-1)}r^{n}_{+}+\frac{2(2n-3)}{(n-1)(n-2)}q^2r^{2-n}_{+}-\frac{3(n-2)}{n-4}r^{n-4}_{+}-\frac{\eta r^{n-2}_{+}}{2\al(n-1)(r^2+d^2)}\left(\left(\frac{\al}{\eta}+\L\right)\times\right.\right.\\\left.\left.r^2_{+}+\frac{q^2}{r^{2(n-2)}_{+}}+\frac{(n-1)(n-2)\bar{q}^2}{2r^{2}_{+}}\right)^2+\frac{\eta}{2\al}\left[\left(\frac{\al}{\eta}+\L\right)^2\left(\sum^{\frac{n-1}{2}}_{j=0}(-1)^jd^{2j}\frac{r^{n-2j}_{+}}{n-2j}+(-1)^{\frac{n-1}{2}}d^n\arctan{\left(\frac{r_+}{d}\right)}\right)+\right.\right.\\\left.\left.2\left(\frac{\al}{\eta}+\L\right)q^2\left(\sum^{\frac{n-5}{2}}_{j=0}(-1)^j\frac{r^{4+2j-n}_{+}}{(4+2j-n)d^{2(j+1)}}+\frac{(-1)^{\frac{n-3}{2}}}{d^{n-2}}\arctan{\left(\frac{r_+}{d}\right)}\right)+q^4\left(\sum^{\frac{3n-7}{2}}_{j=0}(-1)^j\frac{r^{6+2j-3n}_{+}}{(6+2j-3n)d^{2(j+1)}}\right.\right.\right.\\\left.\left.\left.+\frac{(-1)^{\frac{3n-5}{2}}}{d^{3n-4}}\arctan{\left(\frac{r_+}{d}\right)}\right)+(n-1)(n-2)\bar{q}^2\left(\left(\frac{\al}{\eta}+\L\right)\left(\sum^{\frac{n-5}{2}}_{j=0}(-1)^jd^{2j}\frac{r^{n-2j-4}_{+}}{n-2j-4}+\right.\right.\right.\right.\\\left.\left.\left.\left.(-1)^{\frac{n-3}{2}}d^{n-4}\arctan{\left(\frac{r_+}{d}\right)}\right)+q^2\left(\sum^{\frac{n-1}{2}}_{j=0}(-1)^j\frac{r^{2j-n}_{+}}{(2j-n)d^{2(j+1)}}+\frac{(-1)^{\frac{n+1}{2}}}{d^{n+2}}\arctan{\left(\frac{r_+}{d}\right)}\right)\right.\right.\right.\\\left.\left.\left.+\frac{(n-1)(n-2)}{2}\bar{q}^2\left(\sum^{\frac{5-n}{2}}_{j=0}(-1)^j\frac{r^{n+2j-6}_{+}}{(n+2j-6)d^{2(j+1)}}+\frac{(-1)^{\frac{7-n}{2}}}{d^{8-n}}\arctan{\left(\frac{r_+}{d}\right)}\right)\right)\right]\right)
\end{multline}
Similarly as above we give explicit relation for $n=3$ because of a special interest in this case. 
\begin{multline}\label{Gibbs_n3}
G=\frac{1}{4}\left(r_{+}+\frac{\L}{3}r^3_{+}+3\frac{q^2+\bar{q}^2}{r_{+}}+\frac{\eta}{2\al}\left[\left(\frac{\al}{\eta}+\L\right)^2\left(\frac{r^3_{+}}{3}-r_{+}d^2\right)+\frac{(q^2+\bar{q}^2)^2}{r_{+}d^2}\left(\frac{1}{d^2}-\frac{1}{3r^2_{+}}\right)\right.\right.\\\left.\left.+\frac{1}{d}\left(\left(\frac{\al}{\eta}+\L\right)d^2+\frac{q^2+\bar{q}^2}{d^2}\right)^2\arctan{\left(\frac{r_{+}}{d}\right)}-\frac{r_{+}}{2(r^2_{+}+d^2)}\left(\left(\frac{\al}{\eta}+\L\right)r^2_{+}+\frac{q^2+\bar{q}^2}{r^2_{+}}\right)^2\right]\right).
\end{multline}
Since the Gibbs free energy $G$ (\ref{Gibbs_nod}) and its particular case (\ref{Gibbs_n3}) have rather intricate from and their temperature dependences are given implicitly it is difficult to analyse their behaviour. To understand it better we give corresponding graph which shows the dependence $G=G(T)$ while the pressure and all the other parameters are fixed. Namely, the Figure~[\ref{Gibbs_2d}] shows that for smaller pressure $P$ the Gibbs free energy has swallow-tail behaviour and it gives rise to the conclusion that there is a phase transition of the first order and from the qualitative point of view the behaviour of the Gibbs free energy is the same as for Reissner-Nordstrom-AdS black hole in General Relativity \cite{Kubiznak_JHEP12}. The Gibbs free energy in Horndeski gravity for a nonlinearly charged black hole was also examined in our earlier paper \cite{Stetsko_PRD19} and again from the qualitative point of view there is complete agreement between current and earlier results. If the pressure goes up, the swallow-tail gradually diminishes and after reaching of a critical value it completely vanishes and the Gibbs free energy turns to be a smooth function of the temperature $T$ and it also means the disappearance of the phase transition. The critical point when the behaviour of the Gibbs free energy becomes smooth is supposed to be a point of the second order phase transition which is usually takes place for van der Waals system or Reissner-Nordstrom-AdS black hole \cite{Kubiznak_JHEP12}. Due to the interest to the critical point and near critical behaviour some aspects of this issue will be examined in the following section. For better illustration of the swallow-tail behaviour and its gradual diminishing with increasing of the pressure we add the $3D$ figure for the Gibbs free energy (Figure~[\ref{Gibbs_3d}]).

\section{Critical behavior in the extended phase space}
Since additional thermodynamic variables are defined, we are able to extend corresponding thermodynamic phase space for the system and consequently to derive and examine richer thermal behaviour of the black hole. One of the key relations in thermodynamics of any conventional system is its thermal equation of state, which establishes a relation between its macroscopic values such as temperature $T$, pressure $P$ and volume $V$. Having defined the pressure $P$ (\ref{press}) and using the relation for the temperature (\ref{temp}) we can rewrite the latter relation in a form of the thermal equation of state, namely we write:
\begin{equation}\label{eos_1}
P=\frac{1}{8\pi}\left(\frac{q^2}{r^{2(n-1)}_{+}}+\frac{(n-1)(n-2)\bar{q}^2}{2r^4_{+}}-\frac{(n-1)(n-2)}{r^2_{+}}-\xi\right)\pm\frac{1}{4\pi r^2_{+}}\sqrt{2(n-1)\pi\xi r_{+}(r^2_{+}+d^2)T},
\end{equation}
where for convenience we denote $\xi=\al/\eta$ which is directly related to the introduced above thermodynamic value $\Pi$ (\ref{add_var}). We also point out that to obtain the expression (\ref{eos_1}) we extract the cosmological constant $\L$ from the relation (\ref{temp}) solving corresponding quadratic equation for the parameter $\L$ therefore the sign $\pm$ appear in the upper relation. To have the pressure $P$ positive in all the range of variation we pick up the sign $+$ only and consider it in the following relations. We also point out that instead of thermodynamic volume (\ref{TD_vol}) we still keep the horizon radius $r_{+}$, partially because complexity of the relation (\ref{TD_vol}) which does not allows to express $r_+$ as an explicit function of $V$. On the other hand it does not change or modify conclusions about the critical behaviour that we are to derive. In addition we remark that the equation (\ref{eos_1}) being completely ``geometrical" in nature can be rewritten in terms of ``physical" variables in a similar fashion as it was done in \cite{Kubiznak_JHEP12,Gunasekaran_JHEP12}, but such a redefinition of thermodynamic values does not affect on any physical conclusions at all. We also point out that some hints about possible critical behaviour of a nonlinearly charged back hole were obtained in our earlier paper \cite{Stetsko_PRD19}, a more detailed consideration of criticality issues were made in \cite{Hu_PRD19}.

Following the key assumption that the equation of state for black holes (\ref{eos_1}) is analogous to the van der Waals equation of state far reaching consequences can be derived. In particular, critical behaviour can be studied and one of the most important issues here is a phase transition between the so called large and small black holes. The central notion here is the so-called inflection point,  defined as follows:
\begin{equation}\label{infl_eq}
\left(\frac{\partial P}{\partial r_{+}}\right)_{T}=0, \quad \left(\frac{\partial^2 P}{\partial r^2_{+}}\right)_{T}=0.
\end{equation}
It is worth noting that if we use the volume $V$ (\ref{TD_vol}), to find the inflection point the derivatives with respect to the volume $V$ should be equated to zero, but using the relation $\frac{\partial P}{\partial V}=\frac{\partial P}{\partial r_+}\frac{\partial r_+}{\partial V}$, and assuming that the derivative $\frac{\partial V}{\partial r_+}\neq 0$ since the volume is supposed to be a monotonous function of $r_+$ we again arrive at the relations (\ref{infl_eq}).  We also point out that other thermodynamic parameters we used in the extended description are held fixed. The relation for critical radius can be derived straightforwardly using the relations (\ref{infl_eq}), namely after simple calculations we write:
\begin{multline}\label{eq_rc}
-3(n-2)+\frac{(2n-1)q^2}{r^{2(n-2)}_{c}}+\frac{5(n-2)\bar{q}^2}{r^2_{c}}+\frac{3r^4_{c}+22d^2r^2_{c}+15d^4}{2(r^2_{c}+d^2)(r^2_{c}+3d^2)}\left(n-2-\frac{q^2}{r^{2(n-2)}_{c}}-\frac{(n-2)\bar{q}^2}{r^2_{c}}\right)=0,
\end{multline}
where $r_c$ is the critical radius $r_c$. The critical temperature $T_c$ and pressure $P_c$ can be written as functions of the critical radius $r_{c}$:
\begin{equation}\label{cr_T}
T_{c}=\frac{2(n-1)(r^2_c+d^2)}{\pi\xi r_{c}(r^2_{c}+3d^2)^2}\left(n-2-\frac{q^2}{r^{2(n-2)}_{c}}-\frac{(n-2)\bar{q}^2}{r^2_{c}}\right)^2;
\end{equation}
\begin{equation}\label{cr_P}
P_{c}=\frac{1}{8\pi}\left(\frac{q^2}{r^{2(n-1)}_{c}}+\frac{(n-1)(n-2)\bar{q}^2}{2r^4_{c}}-\frac{(n-1)(n-2)}{r^2_{c}}-\xi+\frac{4(n-1)(r^2_{c}+d^2)}{r^2_{c}(r^2_{c}+3d^2)}\left(n-2-\frac{q^2}{r^{2(n-2)}_{c}}-\frac{(n-2)\bar{q}^2}{r^2_{c}}\right)\right).
\end{equation}
The equation for critical horizon radius $r_{c}$ (\ref{eq_rc}) does not have an exact analytical solution for general dimension $n$ and arbitrary chosen parameters $q$, $\bar{q}$ and $d$, therefore the critical values such as $T_{c}$ and $P_c$ cannot be given as explicit functions of the mentioned parameters of the black hole in general case, as it takes for the van der Waals gas or even simpler black hole solutions such as for instance the Reissner-Nordstrom-AdS one \cite{Kubiznak_JHEP12}. In general the critical values can be calculated numerically for arbitrary values of $n$, $\xi$ and black hole charges $q$ and $\bar{q}$. It should be pointed out that for some particular cases analytical solutions can be in principle obtained. Because of some interest in analytical solution and taking into account the fact that analytical solutions often are easier to analyse, we note several particular cases where at least it is possible to derive analytical solution for the critical radius $r_c$ and consequently to other two critical values $T_c$ and $P_c$. First of all, if $n=3$ the equation (\ref{eq_rc}) takes the form:
\begin{equation}\label{eq_rc_n3}
-3+\frac{5(q^2+\bar{q}^2)}{r^{2}_{c}}+\frac{3r^4_{c}+22d^2r^2_{c}+15d^4}{2(r^2_{c}+d^2)(r^2_{c}+3d^2)}\left(1-\frac{q^2+\bar{q}^2}{r^2_{c}}\right)=0.
\end{equation}
The latter equation can be rewritten in a form of a cubic equation for the square of the critical radius $r^2_c$.  Similar equation can be written if the electric charge charge $q=0$, but in this case for any $n$, the only difference with the equation (\ref{eq_rc_n3}) is hidden in the parameter $d$ which is dimension dependent. 

Other interesting particular case is $\al=0$ and it is easy to verify that equation for the critical radius $r_c$ (\ref{eq_rc}) reduces to the form:
\begin{equation}\label{rc_a0}
\frac{(4n-7)q^2}{r^{2(n-2)}_{c}}+5\frac{(n-2)\bar{q}^2}{r^2_c}+2-n=0.
\end{equation}
Corresponding relations for the crictical temperature $T_c$ and the pressure $P_c$ can be rewritten as follows:
\begin{gather}
\label{Tc_a0}T_c=\frac{4(n-2)}{9\pi r_c}\left(1-\frac{\bar{q}^2}{r^2_{c}}-\frac{q^2}{(n-2)r^{2(n-2)}_{c}}\right)^2,\\\label{Pc_a0}P_{c}=\frac{(n-1)(n-2)}{24\pi r^2_{c}}\left(1-\frac{5\bar{q}^2}{2r^2_{c}}-\frac{(4n-7)}{(n-1)(n-2)}\frac{q^2}{r^{2(n-2)}_{c}}\right).
\end{gather}
If $n=3$ the equation (\ref{rc_a0}) turns to be a quadartic one and the crictical radius can be easily written:
\begin{equation}
r^2_{c}=5(q^2+\bar{q}^2).
\end{equation}
Substituting the critical radius into the upper relations (\ref{Tc_a0}) and (\ref{Pc_a0}) we obtain corresponding critical values $T_{c}$ and $P_{c}$ and after the computation we write the explicit expression for the so-called critical ratio:
\begin{equation}\label{rho_a0n3}
\rho_{c}\equiv\frac{P_{c}r_{c}}{T_{c}}=\frac{75}{512}.
\end{equation}
Thus the critical ratio $r_{c}$ as it is expected is a dimesnionless number which does not depend on the parameters of the solution such as its charges $q$, $\bar{q}$, this conclusion is in perfect agreement with the definition of critical ratio for conventional systems as well as in within the extended phase space thermodynamics for black holes. On the other hand it is known that for the standard van der Waals system and the Reissner-Nordstrom-AdS black hole the critical ratio is $\rho_{c}=3/8$ and as we see in our case it is considerably smaller. We also note that exact analytical solutions of the euation (\ref{rc_a0}) can be also derived for $n=4$ and $n=5$, where the equation (\ref{rc_a0}) for $r^2_{c}$ turns to be a quadratic and cubic respectively, but here we do not give explicit relations for corresponding values.

Other important particular cases of the equation (\ref{rc_a0}) are related to the situation when one of the charges are set to zero. Namely, if $q=0$, then the square of the critical radius for any dimension is:
\begin{equation}
r^2_{c}=5q^2.
\end{equation}
Using this result we write the critical ratio $\rho_c$ for this particular case:
\begin{equation}
\rho_c=\frac{75(n-1)}{1024}.
\end{equation}
The obtained relation is in perfect agreement with the relation (\ref{rho_a0n3}) if $n=3$. Finally, we assume that $\bar{q}=0$, then the equation (\ref{rc_a0}) immediately gives us:
\begin{equation}
r^{2(n-2)}_{c}=\frac{(4n-7)}{(n-2)}q^2.
\end{equation}
The latter expression gives rise to the following critical ratio:
\begin{equation}\label{rho_cq0}
\rho_{c}=\frac{3(4n-7)^2}{512(n-2)}.
\end{equation}
Similarly to the upper case there is perfect agreement with the ratio $\rho_{c}$ (\ref{rho_a0n3}) if $n=3$, but in contrast to the upper case its dimension dependence is different. The latter relation also shows that for higher dimensional cases, at least when $n$ is not too high the critical ratio (\ref{rho_cq0}) is also smaller than the corresponding ratio for higher dimensional generalization of the Reissner-Nordstrom-AdS black hole, which equals to $\rho_c=(2n-3)/(4(n-1))$.

If a thermodynamic system undergoes a second order phase transition, there are universal parameters, namely the critical exponents which characterize behaviour of certain thermodynamic values near the critical point and do not depend on the parameters of the system \cite{Goldenfeld_2019}. To obtain the critical exponents it is useful to introduce the so called reduced variables, which show how close to the critical point the system is:
\begin{equation}\label{red_var}
t=\frac{T}{T_{c}}-1, \quad \omega=\frac{r_{+}}{r_{c}}-1.
\end{equation}
Now the critical exponents $\bar{\al}$, $\beta$,$\gamma$ and $\delta$ are defined as follows:
\begin{gather}
C_{V}\sim |t|^{-\bar{\al}},\quad \Delta V_{ls}\sim |t|^{\beta}, \quad \kappa_{T}\sim t^{\gamma}, \quad P-P_c\sim |\omega|^{\delta}.
\end{gather}
Here we point out that $C_V$ is the heat capacity under constant volume, $\Delta V_{ls}$ is the volume difference for large and small phases and $\kappa_T$ is the isothermal compressibility. We also note that instead of the commonly used notation $\al$ for the first of the critical exponents we use $
\bar{\al}$, because the symbol $\al$ is used to denote one of the coupling constants.

It follows from the definition of the entropy $S$ (\ref{entropy}) that the heat capacity under fixed volume exactly equals to zero: $C_V=T\left(\partial S/\partial T\right)_{V}=0$, therefore we immediately conclue that the critical exponent $\al=0$. To derive the other critical exponents we rewrite the equation of state (\ref{eos_1}) near the critical point in the following form:
\begin{equation}\label{eos_cr}
P=P_{c}+At+Bt\omega+C\omega^3+Dt^2+\cdots,
\end{equation}
where:
\begin{gather}
A=T_{c}\left(\frac{\partial P}{\partial T}\right)_{r_{+}}\Big|_{r_c}, \quad B=r_{c}T_{c}\left(\frac{\partial^2 P}{\partial T\partial r_{+}}\right)\Big|_{r_c}, \quad C=\frac{r^3_{c}}{6}\left(\frac{\partial^3 P}{\partial r^3_{+}}\right)_{T}\Big|_{r_c},\quad D=\frac{T^2_{c}}{2}\left(\frac{\partial^2 P}{\partial T^2}\right)_{r_{+}}\Big|_{r_c}.
\end{gather}
The derivatives noted above can be either calculated numerically for a general case of solution or for some particluar cases even analytical expressions can be derived, but in any case the following procedure is identical. Differentiating the equation (\ref{eos_cr}) and taking into account the Maxwell's area law we can write:
\begin{equation}
\int^{\omega_s}_{\omega_l}\omega dP=\int^{\omega_s}_{\omega_l}(Bt+C\om^3)d\om=0.
\end{equation}
After integration we arrive at the relation:
\begin{equation}
Bt(\om^2_{s}-\om^2_{l})+\frac{C}{2}\left(\om^4_{s}-\om^4_{l}\right)=0.
\end{equation}
The obtained equation gives rise to a nontrivial solution $\om_s=-\om_l$. Since for both phases we have the same pressure and using the equation of state (\ref{eos_cr}) we obtain:
\begin{equation}
Bt(\om_s-\om_l)+C\left(\om^3_{s}-\om^3_{l}\right)=0.
\end{equation}
Solving the latter equation for $\om_s$ and taking into account the relation for $\om_s$ and $\om_l$ finally we arrive at the following expression:
\begin{equation}
\om_l\simeq\sqrt{-\frac{B}{C}t}=\sqrt{\frac{B}{C}\frac{(T_c-T)}{T_c}}.
\end{equation}
Now we are able to write the expression for the volume difference $\Delta V_{ls}$ and extract critical exponent from it:
\begin{equation}
\Delta V_{ls}\simeq V_{c}\left(\om_{l}-\om_{s}\right)=2V_{c}\om_{l}\sim |-t|^{1/2}\quad \Rightarrow \quad \beta=\frac{1}{2}.
\end{equation}
Using the definition of the isothermal compressibility $\kappa_{T}$ and the equation of state (\ref{eos_cr}) we can derive the critical exponent $\gamma$
\begin{equation}
\kappa_{T}=-\frac{1}{V}\left(\frac{\partial V}{\partial P}\right)_{T}\sim \frac{1}{Bt}, \quad \Rightarrow \quad \gamma=1.
\end{equation}
Finally, considering the critical isotherm we obtain the critical exponent $\delta$. Namely, from the equation of state (\ref{eos_cr}) it follows:
\begin{equation}
P-P_{c}\sim C\omega^3, \quad \Rightarrow \quad \delta=3.
\end{equation}
All the critical exponents we have derived take the same value as their counterparts for RN-AdS black hole \cite{Kubiznak_JHEP12} and in a case of Horndeski gravity they were derived in the work \cite{Hu_PRD19}, but for a different black hole solution. The same critical exponents were derived for various  solutions in different frameworks, it was mentioned the reviewing paper \cite{Kubiznak_CQG2017}, therefore we can conclude that the critical behaviour shows some universal features, at least for the vast number of black hole solutions in   various independent frameworks. 

We also note that in \cite{Hu_PRD19} the authors used a different equation of state identifying the thermodynamic pressure $P$ not with the cosmological constant $\Lambda$, but relating the pressure with the ratio of the coupling constants $\al/\eta$. In their case that definition of pressure was reasonable, since asymptotic behaviour of the metric function $U(r)$ for infinitely large distances in their case is defined by the ratio $\al/\eta$, in fact that solution has an additional constraint, giving rise to the noted behaviour. In our case we do not impose any specific constraints, thus asymptotic behaviour at the infinity is equally defined by the cosmological constant $\L$  and the ratio $\al/\eta$, actually we have an effective cosmological constant $\Lambda_{eff}\sim \frac{\eta}{\al}\left(\frac{\al}{\eta}-\Lambda\right)^2$, whereas in \cite{Hu_PRD19} the effective cosmological constant is of the form $\L_{eff}\sim\frac{\al}{\eta}$. We also suppose that in our case the thermodynamic pressure can be defined to be proportional to the ratio $\frac{\al}{\eta}$, it will give rise to a bit more cumbersome equation of state instead of equation (\ref{eos_1}), but taking into account the results of the work \cite{Hu_PRD19} we do not think that it changes drastically the critical behaviour or gives rise to other critical exponents.

Since here we focus on the analysis of the thermal behaviour of the system at the critical point or in close vicinity of it we also consider Ehrenfest's equations which are developed for the study of the phase transition of the second order which is supposed to take place at the critical point. The Ehrenfest's equations characterize discontinuity of such thermodynamic parameters as the heat capacity under constant pressure $C_{P}$, the isothermal compressibility $\kappa_{T}$ and the volume expansion coefficient $\tilde{\alpha}$, namely we write:
\begin{gather}\label{ehr_1}
\left(\frac{\partial P}{\partial T}\right)_{S}=\frac{C_{P_2}-C_{P_1}}{VT(\tilde{\al}_2-\tilde{\al}_1)}=\frac{\Delta C_{P}}{VT\Delta\tilde{\al}},\\\label{ehr_2} \left(\frac{\partial P}{\partial T}\right)_{V}=\frac{\tilde{\al}_2-\tilde{\al}_1}{\kappa_{T_2}-\kappa_{T_1}}=\frac{\Delta\tilde{\al}}{\Delta\kappa_T}
\end{gather}
We point here that heat capacity $C_P$ in the upper relation is given by the relation (\ref{heat_capac}), because the latter one was derived under assumption that $\Lambda$ was held fixed. The volume expansion coefficient $\tilde{\alpha}$ is defined as follows: $\tilde{\alpha}=1/V\left(\partial V/\partial T\right)_{P}$. We show that mentioned thermodynamic quanties such as $C_{p}$, $\tilde{\al}$ and $\kappa_{T}$ have infinite discontinuity at the critical point. Let us consider the isothermal compressibility:
\begin{equation}
\kappa_{T}=-\frac{1}{V}\left(\frac{\partial V}{\partial P}\right)_{T}=-\frac{1}{V}\frac{\partial V}{\partial r_{+}}\left(\frac{\partial r_{+}}{\partial P}\right)_{T}.
\end{equation}
Taking into account the first of the conditions (\ref{infl_eq}) we conclude that at the critical point the derivative $(\partial r_{+}/\partial P)_{T}\to\infty$ therefore there is an infinite gap for the isothermal compressibility $\kappa_{T}$ at the critical point. The other two thermodynamic quantities also have an infinite gap at the critical point and it is enough to consider one of them, because for the other one it can be shown in the exactly same way. Let us consider again the heat capacity (\ref{heat_capac}) and it is clear that to show its discontinuity at the critical point we should show that the derivative $(\partial r_+/\partial T)_{P}$ has the infinite gap at the critical point, because both the temperature $T$ and the derivative $\partial S/\partial r_{+}$ are continuous and take finite values at that point. To make analysis more transparent, we write the derivative $(\partial T/\partial r_{+})_{P}$ taken at the critical point $r_c$:
\begin{equation}\label{derT_cr}
\left(\frac{\partial T}{\partial r_{+}}\right)_{P}\Big|_{c}=\frac{r^2_{c}\chi(r_{c})}{8(n-1)\pi\xi(r^2_{c}+d^2)}\left(\frac{(r^2_{c}+3d^2)}{(r^2_{c}+d^2)}\chi(r_c)+2r_{c}\chi'(r_c)\right),
\end{equation}
where we denote $\chi(r)=\xi-\Lambda+(n-1)(n-2)/r^{2}-q^2/r^{2(n-1)}-(n-1)(n-2)\bar{q}^2/2r^4$ and $\chi'(r)$ is its derivative with respect to $r$. Now if we write the derivative $(\partial P/\partial r_{+})_{T}$ at the critical point $r_c$ and  using the expression for the critical temperature $T_c$ (\ref{cr_T}) we obtain:
\begin{equation}\label{derP_cr}
\left(\frac{\partial P}{\partial r_{+}}\right)_{T}\Big|_{c}=-\frac{1}{16\pi r_{c}}\left(\frac{(r^2_{c}+3d^2)}{(r^2_{c}+d^2)}\chi(r_c)+2r_{c}\chi'(r_c)\right)=0.
\end{equation}
Where the last equality is nothing else, but the condition (\ref{infl_eq}), thefore it follows that the expression in the parentheses in the upper relation equals to zero. Since there is identical contribution in the relation (\ref{derT_cr}) we conclude that the derivative $(\partial T/\partial r_{+})_{P}$ equals to zero at the critical point $r_c$ and as a result the heat capacity $C_P$ is discontinuous with infinite gap at this point. 

It is also established that there is a subtlety in the definition of the so-called phase transitions of the second order according to Ehrenfest's classification. Namely, more precisely the character of the phase transition with discontinuous second derivatives as we have here is defined by the Prigogine-Defay ratio, which is introduced as follows:
\begin{equation}\label{PD_ratio}
\tilde{\Pi}=\frac{\left(\partial P/\partial T\right)_{S}}{\left(\partial P/\partial T\right)_{V}}=\frac{\Delta C_{P}\Delta\kappa_{T}}{VT(\Delta\tilde{\alpha})^2},
\end{equation}
obviously the Prigogine-Defay ratio is calculated at the critical point. Taking into account corresponding relations for the thermodynamic values $C_P$, $\tilde{\alpha}$ and $\kappa_{T}$ and substituting them into the upper relation and after simple transformations we obtain:
\begin{equation}
\tilde{\Pi}=-\frac{(\partial S/\partial r_{+})(\partial r_{+}/\partial P)_{T}}{(\partial V/\partial r_{+})(\partial r_{+}/\partial T)_{P}}\Big|_{c}.
\end{equation}
Calculating derivatives $\partial S/\partial r_{+}$ and $\partial V/\partial r_{+}$ and taking into account the relations (\ref{derT_cr}) and (\ref{derP_cr}) we obtain:
\begin{equation}
\tilde{\Pi}=1.
\end{equation}
Therefore, since the Prigogine-Defay equals to unity, the phase transition at the critical point is exactly of the second order. We point out that in contrast to the considered case for dilatonic black holes the Prigogine-Defay ratio is $\tilde{\Pi}<1$ \cite{Stetsko_GRG21} giving rise to the conclusion about a glass-type phase transition for the latter case.

\section{Discussion}
In this work a static charged black hole solution is obtained in Horndeski gravity with linear Maxwell and Yang-Mills fields. Due to chosen form of the field potentials for the gauge fields, namely the Maxwell field is purely electric and the nonabelian field is of magnetic character the explicit relations for the metric function is derived in a closed form. We point out that due to nature of the Horndeski gravity the explicit relations for the metric function $U(r)$ have some differences for even (\ref{U_even_1}) and odd (\ref{U_odd_1}) dimensions of space $n$, this is a specific feature of Horndeski gravity and similar differences occurred even for pure Horndeski gravity \cite{Stetsko_arx18}, but it also affects on the terms related to the gauge field \cite{Stetsko_PRD19}. The other distinctive feature of the obtained solution is a specific ``effective" coupling between the gauge fields which is reflected by the terms proportional to the product of charges $q$ and $\bar{q}$ ($\sim q^2\bar{q}^2$) in the metric function $U(r)$ (\ref{U_mf}) and in the following explicit relations. We point out that  ``effective" coupling of similar character never appears in the framework of General Relativity or Einstein-dilaton theory \cite{Stetsko_GRG21,Stetsko_IJMPA21}, but it may appear in ``higher order" gravity theories for instance when Gauss-Bonnet or higher Lovelock terms are taken into account, but as far as we know it has not been studied yet. It would be interesting to consider this issue in those theories and compare both results. It should be noted that for $n=3$ both abelian and nonabelian fields give identical contribution to the metric function (\ref{U_n3}).

The intricate expression of the metric function $U(r)$ in its integral (\ref{U_mf}) or explicit (\ref{U_odd_1}) (or/and (\ref{U_even_1})) forms turns a thorough analysis of the metric function into a difficult task. But asymptotic cases can be analysed relatively easily. First of all, it follows from (\ref{U_as_inf}) that asymptotic behaviour at the infinity will be of AdS or dS types, depending on the signs of the coupling constants. We also point out that in this case instead of bare cosmological constant $\L$ there is an effective cosmological defined by both the bare constant $\L$ and the ratio of the coupling constants $\al/\e$, namely $\L_{eff}\sim\e/\al\left(\al/\e-\L\right)^2$. It should be noted that imposing additional constraints on the metric functions $U(r)$ and $W(r)$ another effective cosmological constant $\L_{eff}$ can be obtained, namely in \cite{Hu_PRD19} the effective cosmological constant was obtained to be proportional to the ratio of the coupling parameters $\L_{eff}\sim \al/\e$. Therefore it is an interesting issue to examine various options how the effective cosmological constant appears and what form it takes. The latter is also important from the point of view of the extended thermodynamics, because it is directly related to the definition of the thermodynamic pressure $P$. In this work we consider mainly the solution with $AdS$ asymptotic, but as we have mentioned that our solution may have de Sitterian asymptotic depending on the signs of the parameters, but this solution has its own peculiarities and it needs additional careful study.  

For very small distances $r\to 0$ the leading contribution into the metric function $U(r)$ is mainly defined by the gauge field, namely in our case for $n>3$ the dominant contribution is given by the Maxwell field, whereas the for $n=3$ both gauge fields contribute equally. Due to a specific interplay between Horndeski gravity and gauge field terms, the dominant term for $r\to 0$ is always of negative sign, making the behaviour of the metric function more similar to the Schwarzschild solution than to the Reissner-Nordstrom one. In addition the leading term is always proportional to $\sim q^4$ whereas in General Relativity the linear Maxwell field contribution is of the order of $\sim q^2$. The negative sign of the mentioned contribution for $r\to 0$ gives rise to the conclusion that for this particular solution in Horndeski gravity a naked singularity never exist as it might happen for a charged solution in General Relativity, for instance for the Reissner-Nordstrom solution. The Figure~[\ref{metr_f_gr}] confirms the mentioned conclusion. The Figure~[\ref{metr_f_gr}] also shows that increase of the charge (or even both charges) can give rise to appearance of additional horizons, but it needs more careful examination and it will be considered elsewhere.

We also study thermal properties of the black hole. First of all we calculate the black hole temperature, to obtain it we have used the concept of modified surface gravity, introduced in \cite{Hajian_PLB21}, where the authors argued that due to the difference between speeds of gravitons and photons the concept of the surface gravity needs a revision. The modified surface gravity and correspondingly black hole temperature allowed us to avoid introducing additional scalar charges which are ill-defined, as it was done earlier \cite{Feng_PRD16,Stetsko_PRD19} to maintain the first law of black hole thermodynamics. Additional benefit we obtain using the modified surface gravity concept is the fact that the entropy we introduce takes the same form as in General Relativity. To obtain the first law we use the Wald approach \cite{Iyer_PRD94}. We also point out that the concept of the effective surface gravity should be carefully analysed as it is performed in General Relativity. Both temperature $T$ and entropy $S$ allowed us to calculate heat capacity $C_{Q}$ and examine it. Its examination shows that it might have singularity points and instability domains which disappear under certain conditions. These singularities give a hint about possible critical behaviour of the black hole which is also studied in the extended thermodynamics framework.

Finally, introducing the thermodynamic pressure $P$ (\ref{press}) we  obtain the thermal equation of state (\ref{eos_1}). In addition to the pressure we have also introduced the thermal quantity $\Pi$ which has similar nature to the pressure, what was pointed out in \cite{Hu_PRD19}, but this issue should be carefully studied. The extended thermal phase space allowed to derive the Smarr relation (\ref{smarr}). We also obtained the Gibbs free energy. The study of the Gibbs free energy for relatively small pressure shows swallow-tail behaviour (see Figures~[\ref{Gibbs_2d}]-[\ref{Gibbs_3d}]) and increase of the pressure gives rise to gradual diminishing of the swallow-tail behaviour with its following dissolution. The swallow-tail character of  $G=G(T)$ function means that the system undergoes a phase transition of the first order for corresponding values of the pressure $P$ and it disappears when the swallow-tail vanishes with increasing of the pressure. The study of the equation of state (\ref{eos_1}) gives rise to the critical radius $r_c$ (or critical volume $V_c$) which is obtained for some particular cases and consequently for those cases we derived explicit relations for the critical ratios $\rho_{c}$. General relations for the critical values can be studied only numerically. Studying the thermal behaviour near the critical point we obtained the critical exponents $\bar{\al}$, $\beta$, $\gamma$ and $\delta$, their numerical values are the same as even for the Reissner-Nordstrom-AdS black hole \cite{Kubiznak_JHEP12} and for other black hole solution in Horndeski gravity with different equation of state \cite{Hu_PRD19} what confirms universal character of thermodynamic relations. We have also analysed the Ehrenfest's equations to study the behaviour at the critical point and calculated the Prigogine-Defay ratio $\tilde{\Pi}$ which is shown to be equal to one and we make the conclusion that at the critical point we have the second order phase transition. It would also be interesting to study carefully the critical behaviour if instead of the cosmological constant $\L$ the ratio $\al/\eta$ is used to define thermodynamic pressure. The other interesting and important issue is to study in more details the domain where the first-order phase transition occurs, namely to obtain and examine the Clausius-Clapeyron equation. 

\section*{Acknowledgments}
{This was partially supported by the Fulbright Program grant for visiting scholars.}

\end{document}